\documentclass[useAMS,fleqn,usenatbib]{mnras}
\usepackage{ae,aecompl}
\usepackage[T1]{fontenc}

\usepackage{amssymb,amsmath}
\usepackage{wasysym}
\usepackage{color}
\usepackage{longtable}
\usepackage{pdflscape}
\usepackage{mathtools}
\usepackage{float}
\usepackage{footnote}
\makesavenoteenv{tabular}
\makesavenoteenv{table}
\usepackage{epstopdf}
\usepackage{units}
\usepackage{enumerate}

\title[The OCCASO survey: Presentation and radial velocities]
{The OCCASO survey: Presentation and radial velocities of twelve Milky Way Open Clusters\footnote{Based on observations made
with the Nordic Optical Telescope, operated by the Nordic Optical Telescope
Scientific Association, and the Mercator Telescope, operated on the island of La Palma by the Flemish Community,
both at the Observatorio del Roque de los Muchachos,
La Palma, Spain, of the Instituto de Astrof\'isica de Canarias.}
\footnote{Based on observations collected at the Centro Astron\'omico Hispano
Alem\'an (CAHA) at Calar Alto, operated jointly by the Max-Planck Institut f\"ur
Astronomie and the Instituto de Astrof\'isica de Andaluc\'ia (CSIC)}}
\author[Casamiquela et al.]{L. Casamiquela$^{1}$\thanks{E-mail:
lcasam@am.ub.es}, R. Carrera$^{2,3}$, C. Jordi$^{1}$, L. Balaguer-N\'u\~nez$^{1}$, E. Pancino$^{4,5,6}$,\newauthor
S. L. Hidalgo$^{2,3}$,C. E. Mart\'inez-V\'azquez$^{2,3}$, S. Murabito$^{2,3}$, A. del Pino$^{7}$,\newauthor
A. Aparicio$^{2,3}$, S. Blanco-Cuaresma$^{8}$, C. Gallart$^{2,3}$\\ 
$^{1}$Departament d'Astronomia i Meteorologia, Universitat de Barcelona, ICC/IEEC, 08007 Barcelona, Spain\\
$^{2}$Instituto de Astrof\'isica de Canarias, La Laguna, 38205 Tenerife, Spain\\
$^{3}$Departamento de Astrof\'isica, Universidad de La Laguna, 38207 Tenerife, Spain\\
$^{4}$INAF - Osservatorio Astrofisico di Arcetri, Largo Enrico Fermi 5, 50125 Firenze, Italy\\
$^{5}$INAF - Osservatorio Astronomico di Bologna, via Ranzani 1, I-40127 Bologna, Italy\\
$^{6}$ASI Science Data Center, Via del Politecnico SNC, 00133 Roma, Italy\\
$^{7}$Nicolaus Copernicus Astronomical Centre of the Polish Academy of Sciences. ul. Bartycka 18 00-716, Warsaw\\
$^{8}$Observatoire de Gen\`eve, Universit\'e de Gen\`eve, 1290, Versoix, Switzerland\\}

\begin{document}

\date{Accepted 1 March 2016 }

\pagerange{\pageref{firstpage}--\pageref{lastpage}} \pubyear{2016}

\maketitle

\label{firstpage}

\begin{abstract}
Open clusters (OCs) are crucial for studying the formation and evolution of the Galactic disc. However,
the lack of a large number of OCs analyzed homogeneously hampers the investigations about chemical patterns
and the existence of Galactocentric radial and vertical gradients, or an age-metallicity relation.
To overcome this, we have designed the  Open Cluster Chemical Abundances
from Spanish Observatories survey (OCCASO).
We aim to provide homogeneous radial velocities, physical parameters and individual chemical abundances
of six or more Red Clump stars for a sample of 25 old and intermediate-age OCs visible from the Northern hemisphere.
To do so, we use high resolution spectroscopic facilities ($R\geq 62,000$) available at Spanish observatories.
We present the motivation, design and current status of the survey, together with
the first data release of radial velocities for 77 stars in 12 OCs, which represents about 50\%
of the survey. We include clusters never studied with high-resolution spectroscopy before (NGC~1907, NGC~6991, NGC~7762), and
clusters in common with other large spectroscopic surveys like the Gaia-ESO Survey (NGC~6705) and APOGEE (NGC~2682 and NGC~6819).
We perform internal comparisons between instruments to evaluate and correct internal systematics of the results, and compare our 
radial velocities with previous determinations in the literature, when available. Finally, radial velocities for
each cluster are used to perform a preliminar kinematic study in relation with the Galactic disc.
\end{abstract}


\section{Introduction}
Discs are the defining stellar component of most of late-type galaxies, 
including the Milky Way. They contain a substantial fraction of the baryonic 
matter, angular momentum and evolutionary activity of these galaxies, such as 
formation of stars, spiral arms, or bars, and the various forms of 
secular evolution \citep[see][for a review]{vanderKruit+2011}. Understanding 
the formation and evolution of discs is, therefore, one of the key goals of 
galaxy formation research. Two complementary approaches are used to study the growth and 
evolution 
of galactic discs over cosmic time. The first one consists in analyzing discs 
at different redshifts \citep[e.g.][]{Wisnioski+2015}. Although these studies are limited to global 
information integrated over the discs stellar populations, they are able 
to trace the evolution of discs properties with time. The second approach, 
so-called galactic archaeology, consists on reconstructing the disc evolution 
through resolving their stellar populations into individual stars 
\citep[e.g.][]{Carrera+2011a}. The disc evolution is fossilized in the 
orbital distribution of stars, their chemical composition and ages as a function of position: i.e. in 
form of radial and vertical gradients. Part of this information may 
be diluted through dynamical evolution and radial mixing in the disc, which is
less severe for clusters than for field stars. Therefore, the clusters are more
suitable targets for discs studies.

The disc of our own galaxy, the Milky Way, offers an excellent testbed for 
investigating its evolution using all the power of the galactic archaeology 
approach. In spite of the great observational effort performed to unveil the 
details of the disc structure, these are 
still unknown. The vertical density profile has been characterized as 
a sum of two exponential components, the so-called thin and thick discs 
\citep[e.g.][]{Yoshii1982,Gilmore+1983}.  Recent studies have 
focused on dissecting the disc into subsets of stars of very similar chemical 
composition, also called mono-abundance populations 
\citep[e.g.][]{Ivezic+2008}. These studies found that in the solar 
neighborhood the vertical structure is composed of a smooth continuum of disc 
thicknesses \citep[e.g.][]{Bovy+2012}. However, the stellar disc population 
shows a clear bimodal distribution in ([Fe/H],[$\alpha$/Fe]) with two 
sequences of high- and 
low-[$\alpha$/Fe] \citep{Adibekyan+2012,Nidever+2014}. The 
high-[$\alpha$/Fe] is more prominent in the inner disc, while the 
low-[$\alpha$/Fe], and in particular its metal-poor end, dominates in the outer 
disc. \citet{eggen+1962} suggested the 
possibility that the stellar disc formed ``upside-down'' in the sense that old 
stars were formed in a relatively thick component, or are kinematically heated very 
quickly after their birth, while younger populations form in successively 
thinner discs.
It has been thought for a long time that the vertical distribution of the 
disc is the result of some type of heating either due to satellite mergers 
\citep[e.g.][]{Abadi+2003} or radial migration 
\citep[e.g.][]{Sellwood+2002}. However, late results \citep[e.g.][]{Bird+2013} point to an scenario similar 
to the early suggestion by \citet{eggen+1962}.

The radial structure of the Galactic disc has been investigated using different 
tracers trying to cover as much Galactocentric distances as possible. Some of 
these tracers are \ion{H}{ii} regions \citep[e.g.][]{Balser+2011}, B-type 
stars 
\citep[e.g.][]{Daflon+2009}, Cepheid variables 
\citep[e.g.][]{Lemasle+2013,Andrievsky+2013,Korotin+2014,Genovali+2015}, planetary nebulae 
\citep[e.g.][]{Stanghellini+2010}, or open clusters (OCs, see below) and also
main sequence \citep[e.g.][]{Nordstrom+2004,Cheng+2012,Mikolaitis+2014} or 
giant field populations \citep[e.g.][]{Hayden+2014,Huang+2015}. Although all 
of them agree on the existence of a radial metallicity gradient in the sense that stellar populations
are richer towards the inner disk, there are discrepancies about how 
this gradient behaves. While the radial gradient 
described by OCs flattens at large Galactocentric distances 
\citep[e.g.][]{Carrera+2011,Frinchaboy+2013}, the Cepheids do not show 
a slope change in the outer disc \citep[e.g.][]{Lemasle+2013}. These discrepancies can be
partially explained by the fact that each tracer is 
representative of stellar populations of different age.
Until the recent arrival of large 
Galactic surveys, most of the studies were limited by the 
small sample size. The current large surveys are also hampered by the lack of 
accurate distances. This issue will be improved 
significantly in the near future by the advent of \textit{Gaia} space mission data (see Sec.~\ref{sec: gaia}).

In comparison with other tracers, some of the OCs properties, such as distances or 
ages, can be accurately determined \citep[see][for a 
review]{Friel1995}. In fact, most stars, including the Sun, are formed 
in stellar clusters although most of them are dissolved in the first few Myr 
\citep[e.g.][]{Portegies+2010}. Those that survive are the more massive OCs or those
that have had less encounters, which contain the fossil record of the disc formation. 
Moreover, OCs cover a wide range of age 
that allows also to study the evolution of the disc with time 
\citep[e.g.][]{Carrera+2011,Frinchaboy+2013}. The number of clusters old enough ($\gtrsim$ 250 Myr) for such a
study will be increased with \textit{Gaia} observations making this kind of studies even more promising.

For all these reasons, OCs have been used for a long time to investigate the 
Galactic disc, starting from the pioneering studies by 
\citet{Janes1979,Panagia+1980}. A review of the early Galactic 
disc studies using OCs as tracers can be found in \citet{Friel1995}. A 
great observational effort has been performed to characterize OCs homogeneously 
\citep[e.g.][]{Friel+2002,Friel+2010,Sestito+2008,Donati+2015,Bragaglia+2006}
and/or to increase the observed samples 
\citep[e.g.][]{Twarog+1997,Carrera+2011,Jacobson+2011a,Jacobson+2011b}.
All these investigations agree on the fact that the iron content 
decreases with increasing radius as has been found using other tracers 
\citep[e.g.][]{Lemasle+2013}. Most of the previously cited works were limited to the 
inner 15 kpc. However, investigations based on samples containing clusters at 
larger Galactocentric distances \citep[e.g.][]{Carrera+2011,Yong+2012,Frinchaboy+2013} found that the gradient appears 
to flatten from a radius of about 12 kpc, which is near the dynamical
signature for Galactic co-rotation \citep{Lepine+2011}. Moreover, it 
seems that the metallicity gradient observed in the inner disc was steeper in 
the past and has flattened with time 
\citep{Carrera+2011,Jacobson+2011b,Yong+2012,Frinchaboy+2013}, as it is seen in M~33 \citep{Beasley+2015}. No 
significant trends with radius have been observed in the abundances of other 
chemical species \citep[e.g.][]{Yong+2012}. 

\subsection{OCCASO in the context of large surveys}\label{sec: gaia}
Our understanding of the Milky Way in general and the Galactic disc in 
particular is going to change significantly in the next years with the  
\textit{Gaia} space mission 
\citep{Perryman+2001,Mignard2005,Lindegren2005}.
\textit{Gaia} is a full-sky scanning satellite observing all stars down to 20$^{\text{th}}$ magnitude
with precisions at the $\mu$as level. Parallaxes and proper motions of individual stars will be as precise
as 1\% for the OCs up to a distance of 1.5 kpc, and 10\% for almost all known clusters.
Importantly, the faint limiting magnitude and the high precision will allow the discovery
of distant clusters. However, spectroscopic capabilities to derive chemical abundances are limited
due to the low resolution and the small wavelength coverage of the \textit{Gaia} RVS.

On the other hand, the Kepler space mission and its extension K2 is providing asteroseismic data with unprecedented detail,
which will allow to quantify global properties of stars such as age, mass and radii to accuracies near 1\% \citep{Gilliland+2010}.
It is targeting solar-like stars, red giants, classical pulsating stars, and oscillating stars in binaries and clusters.
The advantages of asteroseismology for clusters are that, unlike estimates of colors
and magnitudes, seismic data do not suffer from uncertainties in distance or extinction and reddening.
Asteroseismic observations of many stars allow testing stellar evolution
theory and provide important constraints on the
ages and chemical compositions of stars. K2 data \citep{Howell+2014} is particularly interesting
because it covers a wider area and more clusters than the original Kepler field.

The \textit{Gaia} and Kepler space observations are being complemented with several ongoing and forthcoming
ground-based spectroscopic surveys. Low- and medium-resolution spectroscopic surveys 
($R<10,000$), such as the RAdial Velocity Experiment 
\citep[RAVE;][]{Conrad+2014}, the Sloan Extension for Galactic 
Understanding and Exploration \citep[SEGUE;][]{Lee+2008}, and  Large Sky Area Multi-Object
Fiber Spectroscopic Telescope \citep[LAMOST;][]{Li+2015}
survey, provide radial velocities, together with rough information about the chemical content of 
the studied stars. Large high-resolution spectroscopic surveys 
($R\gtrsim 20,000$) such as the ongoing Apache Point Observatory Galactic Evolution Experiment 
\citep[APOGEE; ][]{Frinchaboy+2013}, the Gaia-ESO Survey 
\citep[GES;][]{Gilmore+2012,Randich+2013}, the GALactic Archaeology with HERMES 
 \citep[GALAH;][]{DeSilva+2015} and the forthcoming WEAVE \citep{Dalton+2012} 
provide detailed information about the chemical composition, in 
addition to radial velocities.

However, most of the large high-resolution spectroscopic surveys do not have dedicated
observations of OCs. Except for a few systems observed for calibration purposes, OCs 
stars are targeted only when they fall in the field of view of other targets. This means 
that the results for most of the studied clusters are based on 
observations of one or two members only. Currently, APOGEE is the only survey sampling the Northern hemisphere. 
GES and GALAH are operating in the South, and WEAVE has not yet defined the observations of OCs and  
will not start operations until at least 2017. APOGEE is obtaining high-resolution 
($R\sim 22,500$) spectra in the infrared $H$-band, which allows to sample the 
innermost regions of the Galaxy. However, it is sampling OC stars at any evolutionary stage and it is not observing a 
minimum of stars in each cluster. In fact, six or more cluster members have been 
analyzed only in 7 of the OCs observed for calibration purposes.
This makes detailed studies of the Milky Way OCs using APOGEE data difficult.

There are other long-term projects dedicated to the study of the OCs. The Bologna Open Cluster Chemical Evolution
project \citep[BOCCE]{Bragaglia+2006} uses both color-magnitude diagram synthesis and
high-resolution spectra to infer cluster properties such as age, distance, and chemical composition.
The WIYN Open cluster study \citep[WOCS]{vonhippel+1998} is also obtaining
photometry, astrometric and spectroscopic data for few nearby OCs. However, these surveys have been designed
to study each cluster individually and not to provide a sample of OCs to
constrain the chemical evolution of the Galactic disc.

Therefore, GES is the only large survey that has a program particularly designed to study
the existence of trends in the Galactic disc. GES is designed to use the FLAMES capabilities
\citep[GIRAFFE$+$UVES;][]{Pasquini+2002} at the second
VLT unit in order to complement the \textit{Gaia} mission. GES clusters observations
include 20-25 OCs older than 0.5 Gyr. For them, GES is using the GIRAFFE fibers 
to derive radial velocities and chemical abundances in stars at any evolutionary stage 
brighter than $V\sim19$ with a resolution $R\sim 20,000$. The six UVES fibers, which cover a 
wavelength range between 4800 and 7000~\AA with a resolution of 47,000, are being used to 
measure accurate radial velocities and detailed chemical abundances
for the brightest targets, mostly Red Clump (RC) stars. The UVES observations
of old OCs have been designed to obtain a homogeneous sample of chemical 
abundances to study the Galactic disc. Using stars in the same evolutionary stage 
avoids the blurring of the trends due to chemical inhomogenities produced by stellar
nucleosynthesis itself, and ensures the homogeneity of the sample.

Several key OCs such as the most metal-rich, NGC~6791, and the oldest, Berkeley~17, 
together with several systems towards the Galactic anticenter or those observed by the Kepler 
mission are only visible from the North, thus will not be observed by GES.

The Open Cluster Chemical Abundances from Spanish 
Observatories (OCCASO) survey has been designed to overcome many of the above caveats. It will obtain 
accurate radial velocities and chemical abundances for more than 20 chemical species from high-resolution spectra 
($R\geq 62,000$) in Northern OCs using the facilities available at Spanish observatories.
As such, it is a natural complement to the GES observations from the South and the 
\textit{Gaia} mission from space. The goal of this paper is to present the survey, its observations,
data reduction, and analysis strategies. We also give a detailed analysis of the radial velocities for the first
batch of observations.
\\
\\
The general survey strategy is described in Sec.~\ref{sec: occaso}.
More in detail: science drivers of the survey (Sec.~\ref{sec: science}) criteria used to select the
cluster sample (Sec.~\ref{sec: sample}), observational facilities used (Sec.~\ref{sec: observations}),
observational strategy (Sec.~\ref{sec: strategy}), and data reduction procedure (\ref{sec: reduction}).
The first data release is described in Sec.~\ref{sec: release}, which includes the description
of the observational material (Sec.~\ref{sec: material}), the accuracy on the wavelength calibration (Sec.~\ref{sec: internal}),
and the results on the radial velocities (Sec.~\ref{sec: results}). Finally, an external comparison
of the stars in common with previous works is done in Sec.~\ref{sec: comparison}, and a discussion of the
results based on the kinematics of the disc and spiral arms are presented in Sec.~\ref{sec: discussion}. A summary is
provided in Sec.~\ref{sec:summary}.

\section{The OCCASO Survey}\label{sec: occaso}

\subsection{OCCASO science drivers}\label{sec: science}
As discussed in the previous section, the main OCCASO science 
driver is the study of the chemical evolution of the Galactic disc. Therefore, 
the observations and analysis strategies have been optimized for this purpose. 
However, the OCCASO 
observational data and results can contribute to 
our understanding of other astrophysical questions. Here we summarize some of 
these additional science topics that can be addressed with OCCASO. 

\begin{itemize}
\item {\sl Galactic disc kinematics.} The same reasons that make OCs 
good chemical tracers of the Galactic disc justify their 
use as tracers to investigate the Galaxy dynamics. The rotation 
curve described by OCs is similar 
to that derived from other thin disc populations such as Cepheids, \ion{H}{ii} 
regions or 
molecular clouds 
\citep[e.g.][]{1987A&A...176...34H,1995AJ....109.1706S,1998A&A...329..514G,
2002AJ....124.2693F}. It seems that the rotational velocity gradually decreases 
with age. This is accompanied by a smooth increase of the line-of-sigh velocity 
dispersion \citep[][]{2014AJ....147...69H}. However, there are several OCs 
with unusual kinematics that keep them away from the disc or the inner regions of 
the Galaxy. It has been suggested that several OCs in the outer disc could have 
been accreted during a dwarf galaxy merger. In this sense, two OCs 
Saurer~1 and Berkeley~29 have been related to the Galactic anticenter stellar 
structure, also known as Monoceros stream \citep{2006AJ....131..922F}. An extragalactic origin has 
also been proposed for the most metal-rich known OC, NGC~6791 
\citep{2006ApJ...643.1151C}. However, accurate proper motions derived from 
Hubble Space Telescope data suggest that this cluster was formed near the Galactic 
bulge \citep{2006A&A...460L..27B}. In addition to the chemical 
abundances OCCASO will provide radial velocities for observed stars with 
uncertainties of about 0.5 km s$^{-1}$ (see Sec.~\ref{sec: results}). These radial 
velocities together with the proper motions provided by the \textit{Gaia} 
mission will allow us to study the three-dimensional kinematics of the OCs,
trace their orbits and relate them to the spiral structure of the Galactic disc.

\item {\sl Stellar evolution laboratories.} OCs have been widely used 
to check the applicability of stellar evolutionary models and the validity of 
their physical parameters and prescriptions such as convective overshooting 
\citep[e.g.][]{2004ApJ...612..168P}, and rotation \citep[e.g.][]{carlberg2014,lanzafame+2015}.
In spite of the progress performed in last 
years, current evolutionary models are not able to completely reproduce the colour-magnitude 
diagrams of many OCs independently of their metallicities 
\citep[e.g.][]{2013MNRAS.430..221A}. A possible explanation could be that each 
cluster has different abundance ratios \citep{2005ARA&A..43..387G}. Stellar 
evolutionary models for different chemical compositions besides the iron 
and $\alpha$-elements have not been available until very recently 
\citep[e.g.][]{2012ApJ...755...15V}. The chemical abundances provided by OCCASO 
will help to constrain the parameters of such.
\end{itemize}

OCCASO could also contribute in the understanding of a variety of topics such as the study of the internal
dynamics of old (highly evolved) OCs \citep[e.g.][]{2003A&A...405..525B,2010ApJ...711..559D},
and the detection of signs of the existence of multiple stellar populations
\citep{2012ApJ...756L..40G,2012ApJ...758..110C,2015ApJ...798L..41C}. However, the small number of stars
sampled in each cluster dificults these kind of studies from OCCASO data only.

\subsection{Clusters and stars selection}\label{sec: sample}
We select OCs to observe in OCCASO according to the following criteria:
\begin{enumerate}[(i)]
 \item Visible from the Northern hemisphere
 \item Ages $\gtrsim0.3$ Gyr, since intermediate-age and old OCs are excellent probes of the
structure and chemo-dynamical evolution of the Galactic disc.
 \item With six or more stars in the expected position of the RC
area of the colour-magnitude diagram (CMD)\footnote{Actually, some bright clusters
not fullfilling this condition were added to be observed during nights of non optimal weather conditions.}.
In general, RC stars are clearly identified even in sparsely populated CMDs.
In some cases, however, it is not easy to differenciate a RC star from a Red Giant Branch (RGB) star in OCs,
so for simplicity we refer them as RC from now on. Selecting RGB stars instead of RC would not imply
abundance changes except maybe for light elements, e.g C or N.
Spectra from these kind of stars are less line-crowded
and therefore, easier to analyze than those of the brighter giants. Moreover, targeting
objects in the same evolutionary state avoids measuring distinct abundances for some elements
due to effects of stellar evolution. The requirement of six stars has
been chosen to have reasonable statistics for the chemical abundances of each cluster.
 \item With RC magnitude brighter than $V \sim 15\,\text{mag}$, constrained by the
available instruments/telescopes.
 \item Prioritizing those with ages, metallicities, heights from the plane, or  
Galactocentric distances lying in poorly studied regions of the $R_{\text{GC}}$-$\left[\text{Fe/H} \right]$,
$Age$-$\left[\text{Fe/H} \right]$, $z$-$\left[\text{Fe/H} \right]$ diagrams.
In this way, we will improve the sampling homogenity of the Galactic disc.
 \item Some clusters with previous high-resolution studies
in the literature \citep[e.g.][]{Carrera+2011,Carrera2012, Bragaglia+2006},
and OCs selected in other surveys (GES, APOGEE) for comparison purposes.
\end{enumerate}

\begin{figure}
\includegraphics[width=0.5\textwidth]{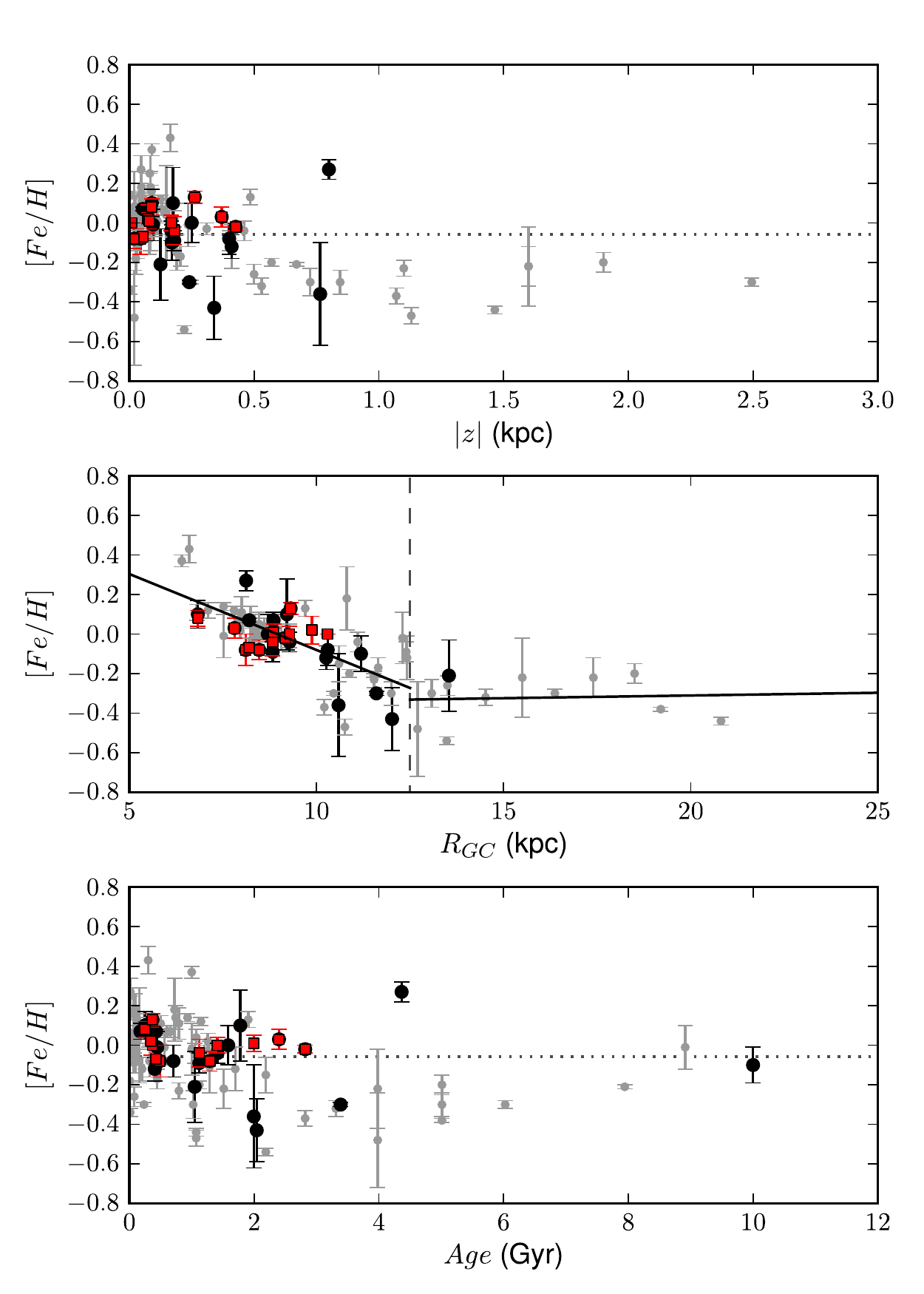}
\caption{[Fe/H] as a function of $|z|$, $R_{\text{GC}}$ and Age. Grey dots correspond to the high-resolution data of OCs compiled
by \citet{Carrera+2011}. Black dots are the full sample of 25 OCs within OCCASO. Red squares are the 12
OCs released in this paper. Solid lines in the middle
panel show the linear fit for OCs inwards and outwards of $R_{\text{GC}} = 12.5\,$kpc.}
\label{fig:trends}
\end{figure}

Following the outlined criteria, we selected a list of 25 candidate OCs,
distributed in the  $R_{\text{GC}}$-$\left[\text{Fe/H} \right]$,
$Age$-$\left[\text{Fe/H} \right]$, $z$-$\left[\text{Fe/H} \right]$ diagrams as seen in Fig.~\ref{fig:trends}. 
This paper focuses on the first 12 OCs for which observations were completed by January 2015. Some basic properties
of these clusters are listed in Table~\ref{clusters}, and they are represented as red squares in Fig.~\ref{fig:trends}. 

\begin{figure*}
\centering
\begin{minipage}{140mm}
\centering
\includegraphics[width=0.7\textwidth]{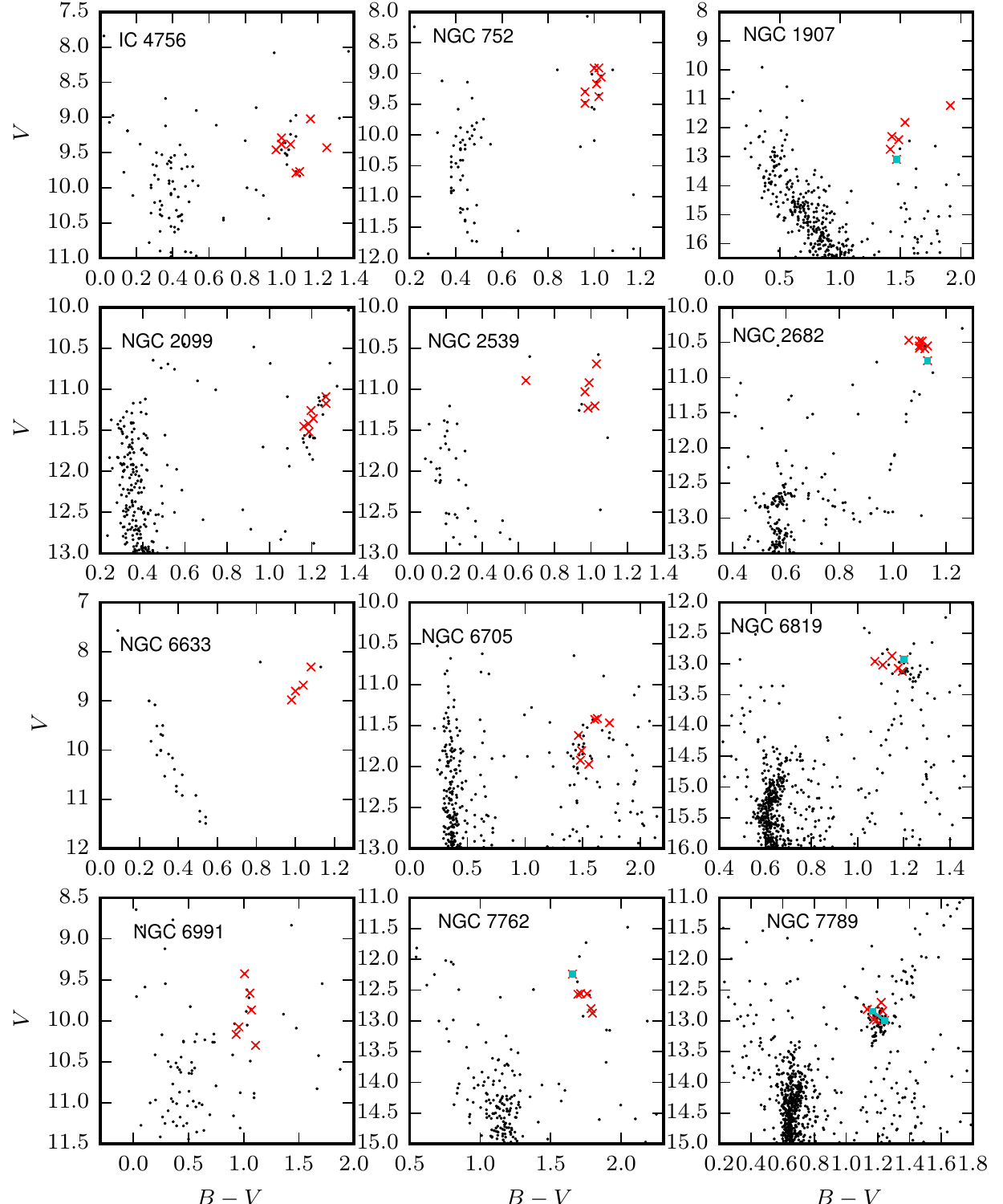}
\caption{(B-V), V colour-magnitude diagrams of the 12 completed clusters (from the photometry listed in
Table~\ref{clusters}). The red crosses indicate the target stars, cyan squares indicate the stars that we have found
to be non members in this study (see Sec.~\ref{final_rv}).}
\label{fig:cmd}
\end{minipage}
\end{figure*}

To select individual stars within each cluster we use the available literature information, with the following procedure:
\begin{enumerate}[(a)]
 \item the targets are first selected among the stars located in the expected position of the RC
in the CMD from the available photometries (see Fig.~\ref{fig:cmd});
 \item membership information based on radial velocities and proper motions, if available,
is taken into account (see Table~\ref{lit});
 \item stars already flagged as non-members or spectroscopic binaries are avoided.
\end{enumerate}

In some cases where membership information is not available (poor photometry, no prior information
about radial velocities or proper motions), we acquire complementary medium-resolution spectroscopy. 
The strategy is to obtain radial velocities and overall metallicities for a large selection of objects in the
line of sight of the cluster, to constrain the selection of members \citep[see][for further details]{Carrera+2015}.

\begin{table}
\begin{centering}
  \caption{\label{clusters}Completed clusters of OCCASO by the end of
January 2015. $D$, $R_{\text{GC}}$, $z$ and Age are from \citet{Dias+2002}.
We list the $V$ magnitude of the RC and the number of stars observed in the last two columns.
The photometry used to select the stars in each OC is indicated as a footnote.}
\begin{tabular}{ccccccc}
\hline 
Cluster & $D$ & $R_{\text{GC}}$ & $z$ & Age & $V_{\text{RC}}$ &Stars \\
 & (kpc) & (kpc) & (pc) & (Gyr) & &  \\
\hline 
IC 4756$^1$ & 0.48 & 8.14 & +41 & 0.50 & 9 & 7 \\
NGC 752$^2$  & 0.46 & 8.80 & -160 & 1.12 & 9 & 7 \\
NGC 1907$^3$  & 1.80 & 10.24 & +9 & 0.31 & 9 & 6 \\
NGC 2099$^4$ & 1.38 & 9.87 & +74 & 0.34 & 12 & 7 \\
NGC 2539$^5$ & 1.36 & 9.37 & +250 & 0.37 & 11 & 6 \\
NGC 2682$^6$ & 0.81 & 9.16 & +426 & 2.81 & 10.5 & 8 \\
NGC 6633$^7$ & 0.38 & 8.20 & +54 & 0.42 & 8.5 & 4$^{\star}$ \\
NGC 6705$^8$ & 1.88 & 6.83 & -90 & 0.25 & 11.5 & 7 \\
NGC 6819$^9$ & 2.51 & 7.81 & +370 & 2.39 & 13 & 6 \\
NGC 6991$^{10}$ & 0.70 & 8.47 & +19 & 1.28 & 10 & 6 \\
NGC 7762$^{11}$ & 0.78 & 8.86 & +79 & 1.99 & 12.5 & 6 \\
NGC 7789$^{12}$ & 1.80 & 9.27 & -168 & 1.41 & 13 & 7 \\
\hline
\end{tabular}
\end{centering}

$^1$\citet{Alcaino1965}; $^2$\citet{Johnson1953}; $^3$\citet{Pandey+2007}; 
$^4$\citet{Kiss+2001}; $^5$\citet{Choo+2003}; $^6$\citet{Montgomery+1993};
$^7$\citet{Harmer+2001}; $^8$\citet{Sung+1999}; $^9$\citet{Rosvick+1998}; 
$^{10}$\citet{Kharachenko+2005}; $^{11}$\citet{Maciejewski+2007};
$^{12}$\citet{Mochejska+1999,McNamara+1981}.
\\
$^{\star}$it has only 4 stars in the RC but was included for observation in a night
with non optimal weather conditions.
\end{table}

\subsection{Observational facilities}\label{sec: observations}
There is no easy access for the European community to a spectrograph with similar 
multi-object capabilities as UVES, in the Northern hemisphere.
However, at Spanish Observatories there are several echelle 
high-resolution spectrographs available with resolutions and wavelength coverage ranges similar 
to, or larger than UVES. In particular, for OCCASO we have selected: CAFE at the 2.2m telescope in the
Centro Astron\'omico Hispano-Alem\'an (CAHA), FIES at the 2.5m NOT telescope in the
Observatorio del Roque de los Muchachos (ORM), and HERMES at the 1.2m Mercator telescope also in the ORM.
See Table~\ref{tab:instr} for a summary of the instrument characteristics.

The high-resolution Fibre-fed Echelle Spectrograph \citep[FIES;][]{Telting+2014} is a cross-dispersed echelle
spectrograph mounted at the 2.5m Nordic Optical Telescope (NOT), and located in the
ORM in the island of La Palma (Spain). FIES is mounted
in a heavily isolated building separated from the NOT building. It is connected to the
Cassegrain focus of the telescope with a fiber bundle offering a maximum resolution
of $R\sim 67,000$. The wavelength coverage of the output spectra is $3700-7300\,\text{\AA}$ without gaps.

The High Efficiency and Resolution Mercator Echelle Spectrograph \citep[HERMES;][]{Raskin+2011} is a fibre-fed
prism-cross-dispersed echelle spectrograph at the 1.2 Mercator telescope, located in the
ORM as well. It is mounted in a temperature-controlled
room and fibre-fed from the Nasmyth A focal station through an atmospheric dispersion corrector.
The size of the detector enables a coverage of the $3770-9000\,\text{\AA}$ wavelength range, with
a maximum resolution of $R\sim 85,000$.

The Calar Alto Fiber-fed Echelle spectrograph \citep[CAFE;][]{Aceituno+2013} is an instrument constructed at the
2.2m telescope in the CAHA in Calar Alto, Almer\'ia (Spain). CAFE
is installed in a temperature and vibration controlled room. It offers a
maximum resolution of $R\sim 62,000$, and a spectral coverage of $3900-9500\,\text{\AA}$.

Since only one star can be observed at once in each of the spectrographs,
we distribute our observations among the three different
telescopes/instruments according to the magnitude of the stars.
This allows us to develop OCCASO on a timeline similar to GES.
The brightest targets ($V\leq 13$) are assigned to HERMES@Mercator, and the
faintest stars ($V> 13$) are assigned mainly to FIES@NOT and CAFE@2.2m CAHA. Current
efficiency of CAFE is lower than expected and all the faint stars were finally moved to FIES.

\begin{table}
 \centering
  \caption{Characteristics of the instruments and telescopes used for the OCCASO Survey.}\label{tab:instr}
  \begin{tabular}{@{}cccc@{}}
  \hline
   Telescope/Instrument & Diameter & Spectral range & Resolution  \\
 \hline
 NOT/FIES & 2.5 m &  $3700-7300\,$\AA & $67,000$ \\
 Mercator/HERMES & 1.2 m & $3770-9000\,$\AA & $85,000$ \\
 2.2mCAHA/CAFE & 2.2 m & $3900-9500\,$\AA & $62,000$ \\
\hline
\end{tabular}
\end{table}

\subsection{Observational strategy}\label{sec: strategy}
All stars are observed in at least 3 exposures lasting 80-3600~s, depending on 
their magnitude, until a global
signal-to-noise ratio (SNR) of at least 70 per pixel at $\lambda \sim6000 
\, \text{\AA}$ is reached. For the faintest targets
($V\geq 14$), this condition is relaxed to a SNR $\sim 50$. Each run we
take a sky exposure to subtract the sky emission lines and, when relevant,
the sky background level (see Sec~\ref{sec: reduction}).
Hot, rapidly rotating stars were
observed twice per run to remove sky absorption features, like telluric bands of 
O$_2$ and H$_2$O. Standard calibration images
(flat, bias and arcs) were also taken at the beginning and end of each night. 
In general we assign each cluster to one instrument to maximize the precision in our measurements.
In order to guarantee the homogeneity of our whole sample, at the beginning of the survey we have repeated 
observations of a set of few stars with the three instruments.
Additionally, Arcturus ($\alpha$-Bootes) and
$\mu$-Leonis, two extensively studied stars, part of the \textit{Gaia} Benchmark stars
\citep{Jofre+2014,Blanco+2014,Heiter+2015} and the APOGEE reference stars \citep{Smith+2013}, were observed with the three 
telescopes for the sake of comparison.
We distribute the target stars among the observing runs (see Sec~\ref{sec: material}) taking into 
account their magnitudes, the quality
of the nights and the characteristics of the instruments.

\subsection{Data reduction}\label{sec: reduction}
The first part of the data reduction consists 
in bias subtraction, flat-field normalization, order tracing and extraction,
wavelength calibration and order merge.
This step is performed with the dedicated pipelines for each instrument: HERMESDRS for
HERMES@Mercator \citep{Raskin+2011}, FIESTool for FIES@NOT \citep{Telting+2014}, and the
pipeline developed by J. Ma\'iz-Apell\'aniz for CAFE@2.2m CAHA, and used in \citet{Negueruela+2014}. 
We have checked that the results from the pipelines are appropriate: the spectra are correctly extracted,
calibration in $\lambda$ is realistic and the merging of the orders does not introduce artefacts
and defects in the regions were orders overlap.
The useful range from CAFE spectra is taken as $4500-9000\,\text{\AA}$ to 
avoid saturated telluric lines and other instrumental defects at the red and blue edges. We take the
whole wavelength ranges for HERMES and FIES. 

After these initial steps of reduction, the spectra from the three instruments are
handled in the same way. The established reduction protocol 
consists in: 
\begin{enumerate}[(i)]
 \item Subtraction of sky emission lines using sky exposures. It was only 
applied to those cases where the levels of the sky lines were higher than 3\% of the 
continuum, to avoid adding noise to the spectra.
 \item Normalization by fitting the continuum with a polynomial function
and radial velocity determination of the individual 
spectra  using DAOSPEC \citep{Stetson+2008} (see details in Sec.~\ref{sec: results}).
 \item Correction of telluric features using the IRAF\footnote{IRAF is distributed by the National Optical Astronomy Observatory,
which is operated by the Association of Universities for Research in Astronomy (AURA)
under a cooperative agreement with the National Science Foundation.} task \textit{telluric}. 
To do so we acquire one or two exposures
of a hot, rapidly rotating star (among HR551, HR7235, HR2198, HR8762 or HR3982, taking into account visibility) in 
each run. The strong O$_2$ band around $7600\,\text{\AA}$ in HERMES and CAFE
spectra is saturated and cannot be removed properly.
 \item Heliocentric correction to account for observer's motion is obtained 
with the IRAF task \textit{rvcorrect}.
 \item The accuracy of the wavelength calibration is tested through the measurement 
of the radial velocity of sky emission lines.
For each run, we measure the radial velocities of the skylines: 
$6300.304$, $6363.78$, $6863.95$, $7276.405$, $7913.708$,
$8344.602$, and $8827.096\, \text{\AA}$ when visible, in all sky exposures and/or in 
target star exposures before applying the heliocentric correction.
The obtained offset, if any, is used to correct the individual exposures with the IRAF task \textit{dopcor} (see Sec.~\ref{sec: internal}).
 \item Combination of the single normalized spectra of the same star and telescope. We 
use the IRAF task \textit{scombine} with a median algorithm and a sigma-clipping rejection. This aims 
to reach the maximum SNR for final radial velocity determination
and further abundance analysis. 
 \item Final radial velocity determination and normalization of the combined 
spectra using DAOSPEC.
\end{enumerate}

\begin{figure}
\centering
\includegraphics[width=0.5\textwidth]{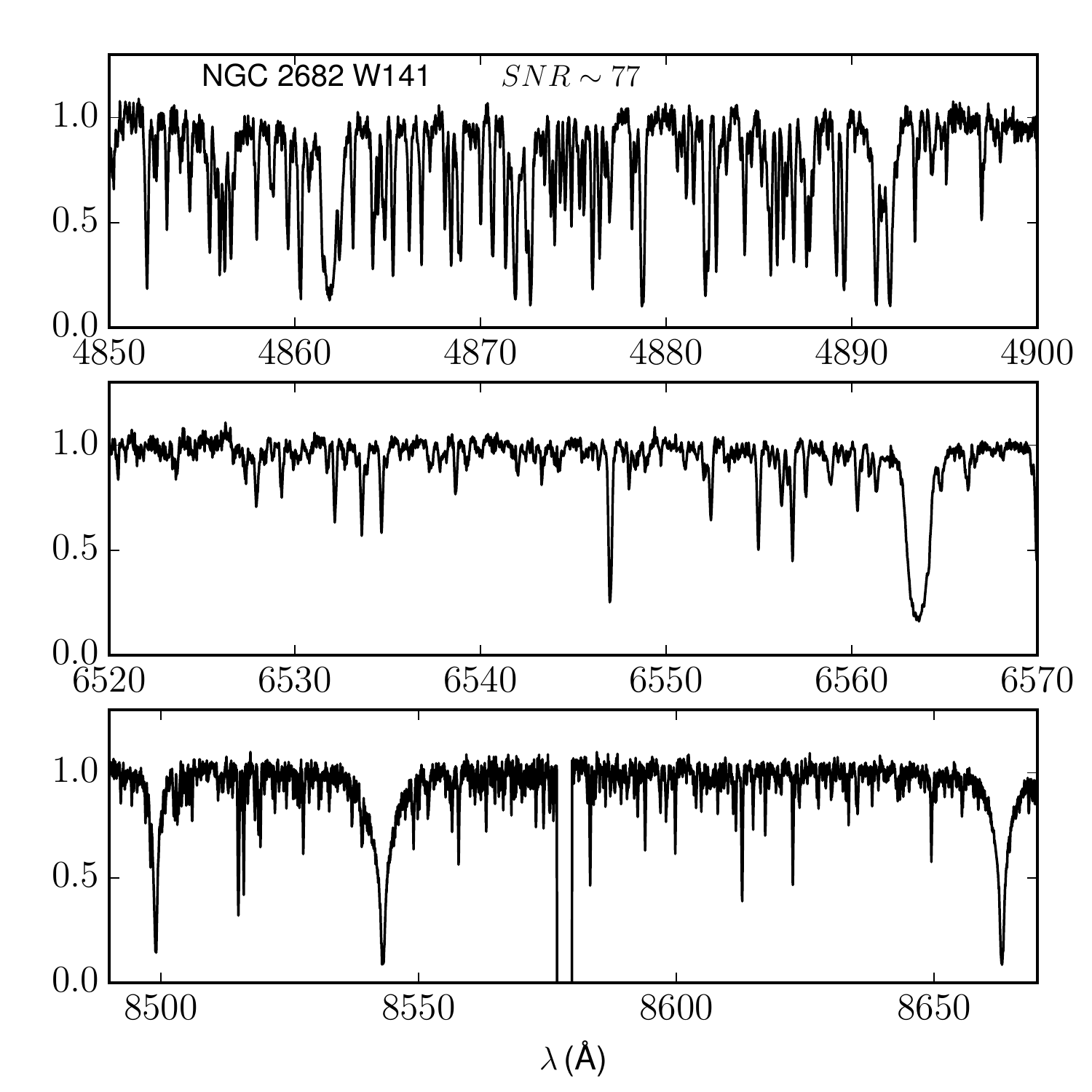}
\caption{The Ca triplet (bottom), H$\alpha$ (middle) and H$\beta$ (top) regions of the final
combined and normalized spectrum of the star NGC~2682 W141 observed with HERMES (SNR$\sim 77$).
A small gap from the order merging can be seen around $8580\, \text{\AA}$.}
\label{fig:spectra}
\end{figure}

As an example of the results of the reduction protocol, we show three regions of the combined
and normalized spectrum of the star NGC~2682 W141 in Fig.~\ref{fig:spectra}.

\section{OCCASO first data release: radial velocities}\label{sec: release}
In this section we present the radial velocities obtained from the reduced spectra acquired
until January 2015 for the completed clusters. 
\subsection{Observational material}\label{sec: material}
OCCASO observations started in January 2013. Until January 2015, we have completed a total
of 53 nights of observations. The number of nights,
dates and instrument of each run are summarized in Table~\ref{runs} together with the 
percentage of time lost due to bad weather, and a description of the quality of the sky.

In this period we have finished observations of 12 clusters which comprise a total of 77 stars (401 spectra), together with
Arcturus and $\mu$Leo used for comparison purposes. For these clusters we have achieved the initial 
requirement of observing at least 6 stars per cluster with a SNR $\sim70$.

\begin{table}
\begin{centering}
\caption{\label{runs}Runs devoted to the project that are included in this paper.}
\begin{tabular}{ccccccc}
\hline 
Run & Period & Instrument & \# & Time & Q$^1$ \\
 & & & nights & lost & \\
\hline 
1 & 1-2 Apr 2013   &FIES  & 2 & 50\% & 2 \\
2 & 25-29 Jul 2013 &HERMES & 5 & 0\% & 1 \\
3 & 23-25 Sep 2013 &FIES & 3 & 50\% & 2 \\
4 & 1-6 Oct 2013   &HERMES & 5 & 30\% & 1 \\
5 & 25-29 Nov 2013 &FIES & 5 & 40\%  & 2 \\
6 & 3-7 Jan 2014 &CAFE & 5 & 100\% & 3 \\
7 & 26 Jan 2014$^2$ &FIES & 1 & 0\% & 2 \\
8 & 29-30 Jan 2014 &CAFE & 2 & 100\% & 3 \\
9 & 21-25 May 2014 &HERMES & 5 & 15\% & 1 \\
10 & 14-15 Jul 2014 &CAFE & 2 & 0\% & 2 \\
11 & 6-8/10-11 Sep 2014 &FIES & 5 & 10\%  & 2 \\
12 & 7-11 Oct 2014 &FIES & 5 & 25\%  & 1 \\
13 & 18-22 Dec 2014 &HERMES & 5 & 15\%& 1 \\
14 & 1-3 Jan 2015 &CAFE & 3 & 0\% & 1 \\
\hline
\end{tabular}
\end{centering}
$^1$Quality of the night:
1: good seeing ($<1''$), no clouds; 2: medium seeing ($1-2''$), disperse thin 
clouds, low dust, we were forced to observe stars 1-2 mag brighter than expected; 3: bad seeing 
($>2''$), clouds, no observations.

$^2$Shared period, only a fraction of the night was used for this project
\end{table}

\subsection{Wavelength calibration accuracy}\label{sec: internal}
The wavelength calibration accuracy is key for the radial velocity determination.
To re-assess it, we calculate the radial velocity offsets of sky emission lines
as described in Sec.~\ref{sec: reduction}.
The mean values and standard deviations of the radial velocity offsets are listed in Table~\ref{skylines}.
We can conclude that:

\begin{enumerate}[(i)]
 \item All FIES runs have negligible offset except for run\#1, for which it has a 
value of $5.09\pm 0.44\, \text{km s}^{-1}$.
The pipeline could not be run in the telescope during the observing run, and it was run
a posteriori using a version built to be used outside the NOT. The origin of the offset could
be related to the use of inappropriate calibration images when running the 
pipeline. We have corrected the individual spectra of this run using this value.
 \item All HERMES offsets are compatible with $0\, \text{km s}^{-1}$ within 
the errors. The mean value is
$-0.28\pm 0.11\, \text{km s}^{-1}$. This offset can be neglected given the 
spectral resolution of the instrument.
 \item Both runs from CAFE present a roughly constant offset of unknown origin, with a mean value and standard
deviation of $2.55\pm0.62\, \text{km s}^{-1}$.
We have shifted all the spectra from these runs by $-2.55\, \text{km s}^{-1}$.
\end{enumerate}

\begin{table}
 \centering
 \caption{\label{skylines}Mean radial velocity offsets and standard deviations for each run 
(number as in Table~\ref{runs}) from visible skylines in the spectra
(see text for more details).}
 \begin{tabular}{cccc}
\hline 
Run & Instrument & $v_{r} \, (\text{km s}^{-1})$ & \# measured lines \\
\hline 
1 &FIES & $5.09\pm0.44$ & 9 \\
3 &FIES & $0.09\pm0.26$ & 5 \\
5 &FIES & $0.07\pm0.24$ & 6 \\
7 &FIES & $-0.04\pm0.17$ & 7 \\
11 &FIES & $-0.5\pm0.7$ & 6 \\
12 &FIES & $0.00\pm0.19$ & 7 \\
\\
2 &HERMES & $-0.16\pm0.28$ & 9 \\
4 &HERMES & $-0.26\pm0.77$ & 7 \\
9 &HERMES & $-0.42\pm0.72$ & 7 \\
13 &HERMES & $-0.29\pm0.89$ & 7 \\
\\
10 &CAFE & $2.45\pm0.52$ & 6 \\
14 &CAFE & $2.64\pm0.72$ & 7 \\
\hline
 \end{tabular}
\end{table}

\subsection{Radial velocities}\label{sec: results}

We present here the results of the radial velocities for stars in the 12 
completed clusters (77 stars), and the reference stars Arcturus and $\mu$-Leo.
This is a total of 79 stars from which 17 have repeated observations with more than one telescope: 
25 were observed with FIES@NOT,
66 were observed with HERMES@Mercator, and 11 were observed with CAFE@2.2m CAHA.

All radial velocities are measured using DAOSPEC \citep{Stetson+2008}.
DAOSPEC is a Fortran code that finds absorption lines in a stellar spectrum, 
fits the continuum, identifies lines from a provided linelist, and
measures equivalent widths. DAOSPEC  also provides radial velocity estimates
using a cross-correlation procedure based on the line centers and on their reference
laboratory wavelength in the linelist (i.e., a sort of line mask cross-correlation).
To run DAOSPEC we used the DOOp code \citep{Cantatgaudin+2014}, an algorithm 
that optimizes its most critical parameters 
in order to obtain the best measurements of equivalent widths (EW). In brief, it fine tunes the 
FWHM and the continuum placement
among other parameters, through a fully automatic and iterative procedure.

We built our linelist starting from the public GES linelist version 3, which
contains 47098 lines. However, this linelist goes from $4700 < \lambda < 6800\, \text{\AA}$
and our covered spectral range is much wider. Therefore, we extended our linelist redder than $6800\, \text{\AA}$
using the linelist described in \cite{Pancino+2010}. The final linelist has 1400 lines, from which
$\sim$1000 (after a sigma clipping rejection criteria) are used for the radial velocities.
Further details will be provided in Casamiquela et al. (in preparation),
where we will release the linelist together with the physical parameters and 
individual abundance determinations from OCCASO.
 
We compute radial velocities from both individual and combined exposures for
each star, as mentioned in Sec.~\ref{sec: reduction}.
Using the combined exposures, we perform a comparison among the three instruments, and we
compute the final values per star. We perform a membership selection after which we compute the
average radial velocity for each of the 12 clusters. Details are given in the following subsections.

\subsubsection{Individual exposures}\label{sec: rv indiv}

We measure radial velocities from individual exposures after rectifying the
offsets calculated in Sec.~\ref{sec: internal},
and once heliocentric corrections are applied. The values obtained are listed 
in Table~\ref{vrindiv}. The first, second and third
columns denote the star identifier (taken from WEBDA\footnote{\url{http://www.univie.ac.at/webda/}}), night of observation, and instrument, 
respectively; the fourth column indicates the Heliocentric Julian Date (HJD) of
the observation; and the fifth column lists the measured radial velocity 
and the uncertainty. The quoted uncertainties are those calculated by DAOSPEC, which correspond to the
line-by-line radial velocity variance.

\begin{figure}
\centering
\includegraphics[width=0.5\textwidth]{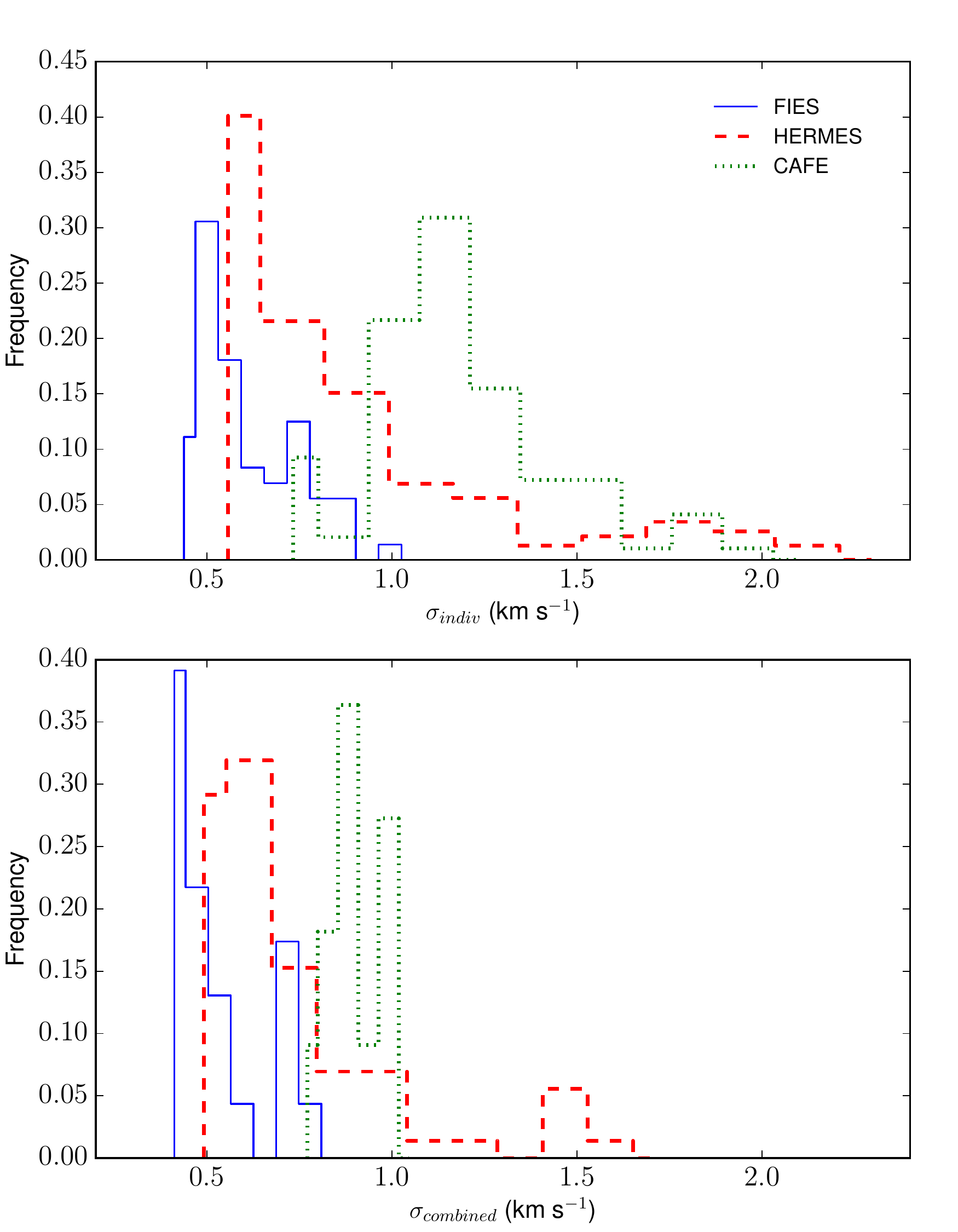}
\caption{Radial velocity uncertainty distributions from the 
individual spectra (top panel), and the combined spectra (bottom
panel), for each instrument. The histograms are scaled to facilitate the visualization.}
\label{fig:sigma}
\end{figure}

The uncertainties on the individual radial velocities are constrained by the 
resolution and wavelength range (which limits the number of lines used)
 of the instrument, and the SNR of the spectrum.
The distribution of uncertainties is shown in Fig.~\ref{fig:sigma}, with median 
values of $0.6\pm 0.1\, \text{km s}^{-1}$ for FIES, $0.8\pm 0.4\, \text{km s}^{-1}$ for 
HERMES, and $1.2\pm 0.3\, \text{km s}^{-1}$ for CAFE.

Although our observations are not designed to look for spectroscopic binaries\footnote{
in many cases several observations are consecutive}, we can detect
them by comparing the radial velocity obtained from different exposures of the same star.
Individual radial velocities for all stars agree within the errors but one,
NGC~6819 W983, with a radial velocity of $3.2\pm 0.8\, \text{km s}^{-1}$ from 
the exposure in the night 25 Jul 2013,
and $-8.3\pm 0.8\, \text{km s}^{-1}$ from the three consecutive exposures in 
the night 29 Jul 2013. We flag this star as possible spectroscopic binary,
(see Sec. 3.3.3 for further discussions).

There can be other single-line spectroscopic binaries within our sample that we are not detecting
because in most cases we have taken the individual
exposures in the same night. In this case we would only detect them if the 
period is very short.

\begin{table}
\scriptsize
\centering
 \caption{\label{vrindiv}Radial velocities from individual spectra. The complete version of the table
can be found in the electronic version of the Journal, and in the CDS.}
 \begin{tabular}{cccccc}
\hline 
Star & Night & Instr & HJD & $v_{\text{r,indiv}}$ \\
 &  &  &  & (km s$^{-1}$) \\
\hline 
IC4756 W0042 & 20130729  & HERMES &  2456503.42986657 &   $-24.7 \pm0.6$   \\
IC4756 W0042 & 20130729  & HERMES &  2456503.4350752  &   $-24.7 \pm0.6$   \\
IC4756 W0042 & 20130729  & HERMES &  2456503.44028436 &   $-24.7 \pm0.6$   \\
IC4756 W0042 & 20140521  & HERMES &  2456799.71796826 &   $-24.5 \pm0.7$   \\
IC4756 W0042 & 20140521  & HERMES &  2456799.72317693 &   $-24.5 \pm0.7$   \\
\hline 
\end{tabular}
\end{table}

\subsubsection{Combined spectra and comparison among instruments} \label{sec: combined}

The final values of the radial velocities are obtained running again DOOp on the 
combined spectra. The results of each star and instrument are specified in
columns 9, 10 and 11 (for FIES, HERMES and CAFE, respectively) of Table~\ref{lit}.
The radial velocity uncertainties are reduced with respect to the ones from
individual spectra due to the higher SNR,
as shown in the lower pannel of Fig.~\ref{fig:sigma}. Now the median dispersion values for each instrument are: 
$0.5\pm 0.1\, \text{km s}^{-1}$ for FIES, $0.7\pm 0.3\, \text{km s}^{-1}$ for 
HERMES, and $0.93\pm 0.07\, \text{km s}^{-1}$ for CAFE.

\begin{figure}
\centering
\includegraphics[width=0.35\textwidth]{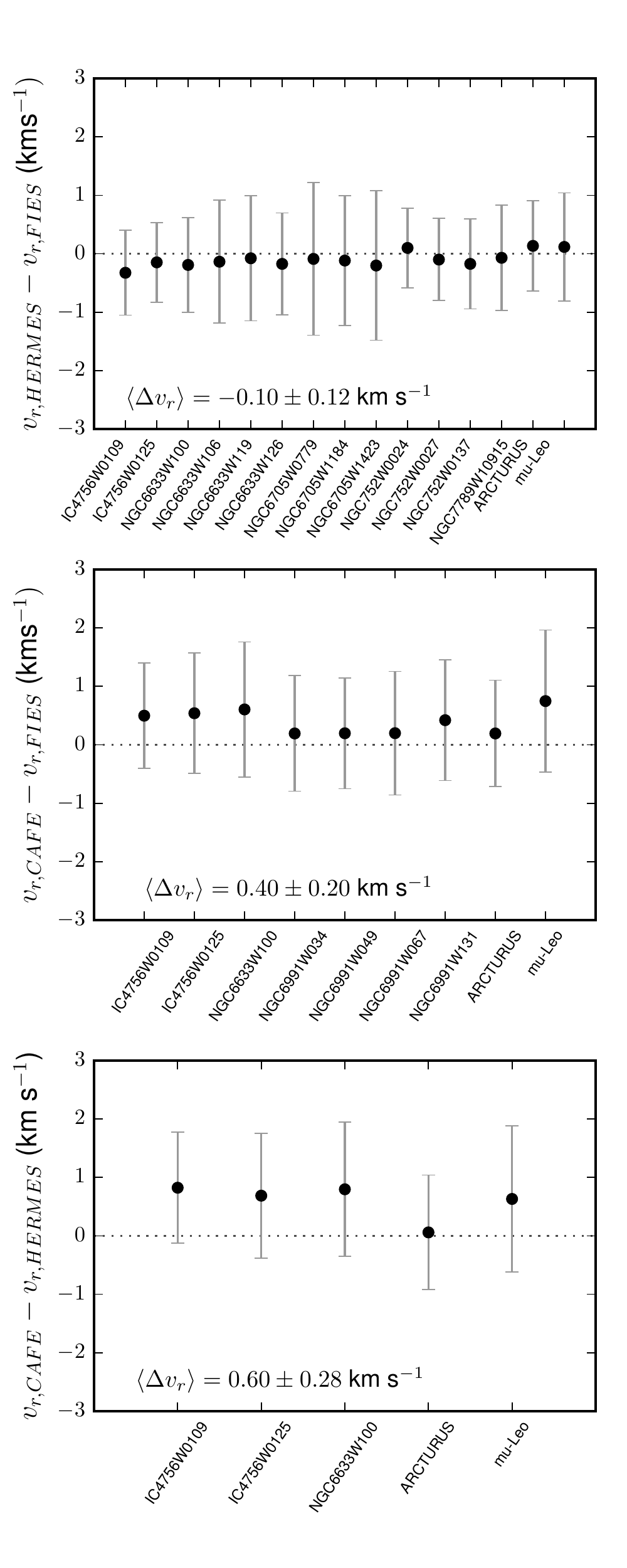}
\caption{Differences in $v_{r}$ obtained for the stars in common between 
HERMES@Mercator and FIES@NOT (top panel), CAFE@2.2m CAHA and FIES@NOT
(central panel), and CAFE@2.2m CAHA and HERMES@Mercator (bottom panel).The error bars are
the sum in quadrature of the two uncertainties.}
\label{fig:comp_tel}
\end{figure}

We use the final combined spectra of the repeated stars to make a comparison 
among instruments (see Fig.~\ref{fig:comp_tel}).
Fifteen stars were observed with both FIES@NOT and HERMES@Mercator,
nine stars observed with both CAFE@2.2m CAHA and FIES@NOT, and five stars 
observed with both HERMES@Mercator and CAFE@2.2m CAHA. We notice:
\begin{enumerate}[(i)]
 \item For HERMES-FIES comparison, we find a mean offset and dispersion of
$\langle \Delta v_\text{r} \rangle=-0.10\pm 0.12\, \text{km s}^{-1}$.
 \item For CAFE-FIES, we find a mean offset of $\langle \Delta v_\text{r} 
\rangle=0.40\pm 0.20\, \text{km s}^{-1}$.
 \item For the CAFE-HERMES case, we find a mean offset of $\langle \Delta v_\text{r} 
\rangle=0.60\pm 0.28\, \text{km s}^{-1}$.
\end{enumerate}
All offsets are in agreement within the observational uncertainties and follow the
expectations from sky emission lines results
(see Table~\ref{skylines}, Sec~\ref{sec: internal}).

\begin{landscape}
\begin{table}
\begin{centering}
\tiny
\caption{Radial velocities obtained with FIES, HERMES and CAFE, and the combination of all instruments $v_{\text{r,OCCASO}}$.
Values from literature are $v_{\text{r,ref}}$, and differences with literature are computed as 
$\Delta v_{\text{r}}=v_{\text{r,OCCASO}}-v_{\text{r,ref}}$. Information on membership in the literature is shown: probability from proper motion
($P_{\mu}$), from radial velocity $P_{v_{\text{r}}}$, and membership classification (Class). Last column points out special cases discussed
in the text. Star IDs are from WEBDA. The complete table can be found in the electronic version of
the Journal, and in the CDS.}\label{lit}
\begin{tabular}{@{}cccccccccccccccccc@{}}
\hline
Cluster & Star & RA & DEC & $V$ & $P_{\mu}$ & $P_{v_{\text{r}}}$ & Class & $v_{\text{r,FIES}}$ & $v_{\text{r,HERMES}}$ & $v_{\text{r,CAFE}}$ & $v_{\text{r,OCCASO}}$ & $ v_{\text{r,ref}}$ & $\Delta v_{\text{r}}$ & Reference &Remark\\
\hline
 & Arcturus & 14:15:39.672 & +19:10:56.67 & -0.05  & & & & $-5.1 \pm 0.5$ &$-5.0 \pm 0.6$ &$-4.9 \pm 0.8$ &$-5.0 \pm 0.9$ &$-5.19 \pm 0.03$ & 0.19&  \cite{Blanco+2014} & \\
 & $\mu$-Leo & 09:52:45.817 & +26:00:25.03 & 3.88  & & & & $13.7 \pm 0.6$ &$13.8 \pm 0.7$ &$14.5 \pm 1.0$ &$13.9 \pm 1.2$ &$13.53 \pm 0.03$ & 0.37&  \cite{Blanco+2014} & \\
\hline
IC 4756 & W0042 & 18:37:20.77 & +05:53:43.1& 9.46& & &  & & $-24.7 \pm 0.6 $ & & $-24.7 \pm 0.6 $ &$-24.9 \pm 0.2$& 0.2 &  \cite{Mermilliod+2008} & \\ 
        &       & & & & & & & & & & & $-25.2 \pm 0.7$& 0.5 &  \cite{Valitova+1990} & \\ 
 	& W0044 & 18:37:29.72 & +05:12:15.5 & 9.79& 0.96$^1$ & & & & $-25.8 \pm 0.7 $ & & $-25.8 \pm 0.7 $ & $-26.0 \pm 0.1$& 0.2 &  \cite{Mermilliod+2008} & \\ 
 	&       & & & & & & & & & & & $-26.6 \pm 0.2$& 0.8 &  \cite{Valitova+1990} & \\ 
 	& W0049 & 18:37:34.22 & +05:28:33.5 & 9.43& 0.96$^1$ & & & & $-25.2 \pm 0.6 $ & & $-25.2 \pm 0.6 $ & $-25.4 \pm 0.1$& 0.2 &  \cite{Mermilliod+2008} & \\ 
 	&       & & & & & & & & & & & $-26.0 \pm 0.4$& 0.8 &  \cite{Valitova+1990} & \\ 
 	& W0081 & 18:38:20.76 & +05:26:02.3 & 9.38& 0.91$^1$,0.99$^{10}$ & & & & $-23.1 \pm 0.7 $ &  & $-23.1 \pm 0.7 $ & $-23.2 \pm 0.1$ & 0.1 &  \cite{Mermilliod+2008} & \\ 
 	&       & & & & & & & & & & & $-27.9 \pm 0.5$ & 4.8 &  \cite{Valitova+1990} & \\ 
 	& W0101 & 18:38:43.79 & +05:14:20.0 & 9.38& 0.94$^1$,0.99$^{10}$ & & & & $-25.5 \pm 0.7 $ & & $-25.5 \pm 0.7 $ & $-25.7 \pm 0.1$&   0.2&  \cite{Mermilliod+2008} &\\ 
 	&       & & & & & & & & & & & $-25.6 \pm 0.2$&   0.1&  \cite{Valitova+1990} &\\ 
 	& W0109 & 18:38:52.93 & +05:20:16.5 & 9.02& 0.96$^1$,0.99$^{10}$ & & & $-24.5 \pm 0.5 $ & $-24.8 \pm 0.6 $ & $-24.0 \pm 0.9 $ & $-24.5 \pm 0.6$ & $-25.2 \pm 0.1$ & 0.7&  \cite{Mermilliod+2008}  &\\ 
 	&       & & & & & & & & & & & $-24.4 \pm 0.4$ & -0.1&  \cite{Valitova+1990} & \\ 
 	& W0125 & 18:39:17.88 & +05:13:48.8 & 9.29& 0.92$^1$,0.99$^{10}$ & & & $-24.5 \pm 0.4 $& $-24.7 \pm 0.5 $ & $-24.0 \pm 0.9 $ &  $-24.5 \pm 0.6$ & $-24.9 \pm 0.1$& 0.4 &  \cite{Mermilliod+2008} &\\ 
 	&       & & & & & & & & & & & $-24.4 \pm 0.4$& -0.1 &  \cite{Valitova+1990} &\\ 
\hline
NGC 752  & W0001 & 01:55:12.60 & +37:50:14.60 &9.48& 0.93$^2$,0.93$^{10}$ & & M$^{\text{a}}$& $5.3 \pm 0.4 $ &  & & $5.3 \pm 0.4 $ & $5.2 \pm 0.1$ & 0.1&  \cite{Mermilliod+2008} & \\ 
 	&       & & & & & & & & & & & $4.73 \pm 0.20 $& 0.57 &  \cite{BocekTopcu+2015} &\\ 
         & W0024 & 01:55:39.35 & +37:52:52.69 & 8.91& 0.99$^2$,0.93$^{10}$& & M$^{\text{a}}$& $5.6 \pm 0.4 $ & $5.7 \pm 0.5 $ & & $5.6 \pm 0.5 $ & $5.4 \pm 0.1$ & 0.2&  \cite{Mermilliod+2008}  &\\ 
 	&       & & & & & & & & & & & $4.86 \pm 0.19$& 0.74 &  \cite{BocekTopcu+2015} &\\ 
         & W0027 & 01:55:42.39 & +37:37:54.66 &9.17& 0.99$^2$,0.93$^{10}$ & & M$^{\text{a}}$& $4.9 \pm 0.5 $ & $4.8 \pm 0.5 $ & & $4.9 \pm 0.5 $& $4.6 \pm 0.1$ & 0.3&  \cite{Mermilliod+2008} & \\ 
 	&       & & & & & & & & & & & $4.39 \pm 0.19$& 0.51 &  \cite{BocekTopcu+2015} &\\ 
         & W0077 & 01:56:21.63 & +37:36:08.53 & 9.38&0.98$^2$,0.91$^{10}$ & & M$^{\text{a}}$& & $5.2 \pm 0.5 $  & & $5.2 \pm 0.5 $& $5.0 \pm 0.1$ & 0.2&  \cite{Mermilliod+2008} & \\ 
 	&       & & & & & & & & & & & $4.58 \pm 0.20$& 0.62 &  \cite{BocekTopcu+2015} &\\ 
         & W0137 & 01:57:03.12 & +38:08:02.73 & 8.90& 0.99$^2$,0.93$^{10}$& & M$^{\text{a}}$& $5.7 \pm 0.5 $ & $5.5 \pm 0.6 $ & & $5.6 \pm 0.5 $& $5.2 \pm 0.1$  & 0.4&  \cite{Mermilliod+2008}  &\\ 
 	&       & & & & & & & & & & & $5.59 \pm 0.20$& 0.01 &  \cite{BocekTopcu+2015} &\\ 
         & W0295 & 01:58:29.81 & +37:51:37.68 & 9.30& 0.99$^2$,0.93$^{10}$& & M$^{\text{a}}$& $5.6 \pm 0.5 $ & & &$5.6 \pm 0.5 $ & $5.2 \pm 0.1$  & 0.4&  \cite{Mermilliod+2008}  &\\ 
 	&       & & & & & & & & & & & $6.32 \pm 0.23 $& -0.72 &  \cite{BocekTopcu+2015} &\\ 
         & W0311 & 01:58:52.90 & +37:48:57.30 & 9.06& 0.99$^2$,0.92$^{10}$& & M$^{\text{a}}$& & $6.0 \pm 0.6 $ & & $6.0 \pm 0.6 $& $5.8 \pm 0.1$ & 0.2 &  \cite{Mermilliod+2008} &\\ 
 	&       & & & & & & & & & & & $5.19 \pm 0.19$& 0.81 &  \cite{BocekTopcu+2015} &\\ 
\hline
NGC 1907 & W0062 & 05:27:49.053 & +35:20:10.13 & 12.41& 0.98$^{10}$& & M$^\text{b}$ & &$ 2.6\pm 1.6$ & &$ 2.6\pm 1.6$&$-2.08 \pm 1.4$ & 4.68&  \cite{Glushkova+1991} & \\
         & W0113 & 05:28:04.207 & +35:19:16.32 & 11.81& 0.61$^{10}$& & M$^\text{b}$ & &$ 2.2\pm 0.6$ & &$ 2.2\pm 0.6$&$1.67 \pm 0.9$  & 0.53&  \cite{Glushkova+1991} & \\ 
         & W0131 & 05:28:05.276 & +35:19:49.64 & 12.30& 0.98$^{10}$& & M$^\text{b}$ & &$ 2.3\pm 1.2$ & &$ 2.3\pm 1.2$& $-0.68 \pm 2$ & 2.98&  \cite{Glushkova+1991}  &\\ 
         & W0133 & 05:28:05.863 & +35:19:38.87 & 12.74& 0.98$^{10}$& & & &$-0.2\pm 1.7$ & &$-0.2\pm 1.7$&  &   & \\ 
         & W0256 & 05:28:01.783 & +35:21:14.89 & 11.23& 0.98$^{10}$& & M$^\text{b}$& &$ 2.8\pm 0.8$ & &$ 2.8\pm 0.8$& $1.45 \pm  0.69$ & 1.35&  \cite{Glushkova+1991}  &\\ 
         & W2087 & 05:27:38.899 & +35:17:18.04 & 13.09& & & & &$63.4\pm 1.0$ & &$63.4\pm 1.0$&  &   & & X \\ 
\hline
NGC 2099 & W007  & 05:52:20.31 & +32:33:49.3 & 11.42& 0.85$^3$,1.00$^{10}$ & & &  & $8.9 \pm 0.9 $ & &$8.9 \pm 0.9 $ &$8.7 \pm 0.2$ & 0.2 &  \cite{Mermilliod+2008} & \\ 
         & W016  & 05:52:17.26 & +32:32:56.5 & 11.26& 0.89$^3$,0.98$^{10}$ & & &  & $7.5 \pm 0.9 $ & &$7.5 \pm 0.9 $ &$7.2 \pm 0.2$ & 0.3 &  \cite{Mermilliod+2008}  &\\ 
 	 & W031  & 05:52:16.68 & +32:31:39.3 & 11.52& 0.87$^3$,0.98$^{10}$ & & &   & $7.4 \pm 1.3 $ & &$7.4 \pm 1.3 $ &$7.1 \pm 0.2$ & 0.3 &  \cite{Mermilliod+2008} & \\ 
 	 & W148  & 05:52:08.10 & +32:30:33.1 & 11.09& 0.86$^3$,0.93$^{10}$& & &   & $8.6 \pm 0.9 $ & &$8.6 \pm 0.9 $ &$8.7 \pm 0.2$ & -0.1 &  \cite{Mermilliod+2008} & \\ 
 	 &       &   & & && & &  & & & & $9.1 \pm 0.4$ & -0.5 &  \cite{Pancino+2010}  &\\ 
 	 & W172  & 05:52:04.89 & +32:33:18.3 & 11.45& 0.79$^3$,0.99$^{10}$& & &   & $8.2 \pm 0.9 $ & &$8.2 \pm 0.9 $ &$8.1 \pm 0.2$ & 0.1 &  \cite{Mermilliod+2008} & \\ 
 	 & W401  & 05:51:55.14 & +32:30:03.0 & 11.36& 0.90$^3$,0.98$^{10}$& & &    & $9.5 \pm 0.7 $ & &  $9.5 \pm 0.7 $&$9.2 \pm 0.2$ & 0.3 &  \cite{Mermilliod+2008} & \\ 
 	 & W488  & 05:52:46.97 & +32:33:19.4 & 11.17& 0.87$^3$,0.97$^{10}$& & &   & $8.6 \pm 0.8 $ & &$8.6 \pm 0.8 $ &$8.3 \pm 0.2$  & 0.3&  \cite{Mermilliod+2008} & \\ 
\hline
NGC 2539 & W229  & 08:10:33.80 & -12:51:48.9 & 11.20& 0.99$^{10}$ & & & &$29.8\pm 0.7$ & &$29.8\pm 0.7$ &  &   \\ 
         & W233  & 08:10:34.35 & -12:49:55.2 & 10.89& 0.99$^{10}$ & & & &$34.8\pm 1.1$ & & $34.8\pm 1.1$ &$26.7 \pm 0.2 $ & 8.1 &  \cite{Mermilliod+2008} & X \\ 
         & W251	 & 08:10:38.99 & -12:44:44.7 & 11.23& 0.98$^{10}$ & & & &$29.4\pm 1.0$ & & $29.4\pm 1.0$ & $29.4 \pm 0.2 $ & 0&  \cite{Mermilliod+2008} & \\
         & W346  & 08:10:23.02 & -12:50:43.3 & 10.92& 0.97$^{10}$ & & & &$30.0\pm 0.6$ & &$30.0\pm 0.6$ & $29.7 \pm 0.1 $ & 0.3 &  \cite{Mermilliod+2008} &\\ 
         &  &  & & &  & & & & & &  & $29.7 \pm 0.2 $ & 0.3 &  \cite{Reddy+2013} &\\ 
         & W463  & 08:10:42.87 & -12:40:11.8 & 10.69&  & & & &$29.0\pm 0.7$ & & $29.0\pm 0.7$& $28.8 \pm 0.1 $ & 0.2 &  \cite{Mermilliod+2008} &\\ 
         &  &  & & &  & & & & & & & $28.7 \pm 0.2 $ & 0.3 &  \cite{Reddy+2013} &\\ 
         & W502  & 08:11:27.67 & -12:41:06.8 & 11.03&  & & & &$28.9\pm 0.8$ & & $28.9\pm 0.8$ & $28.9 \pm 0.2 $ & 0 &  \cite{Mermilliod+2008} &\\ 
\hline
\end{tabular}
\end{centering}
\noindent Membership probabilities: $P_{\mu}$: $^1$\cite{Herzog+1975}, $^2$\cite{Platais1991},
$^3$\cite{Zhao+1985}, $^4$\cite{Sanders1977}, $^5$\cite{Sanders+1973}, $^6$\cite{McNamara+1977}, $^7$\cite{Sanders1972},
$^8$\cite{Kharchenko+2005}, $^9$\cite{McNamara+1981}, $^{10}$\cite{Dias+2014};
$P_{v_{\text{r}}}$: $^\text{i}$\cite{Geller+2015}, $^{\text{ii}}$\cite{Milliman+2014}\\
Membership classification provided by literature from: $^\text{a}$\cite{BocekTopcu+2015}, $^\text{b}$\cite{Geller+2015},
$^\text{c}$\cite{cantat-gaudin+2014},
$^\text{d}$\cite{Mathieu+1986},$^{\text{e}}$\cite{Milliman+2014}, $^\text{f}$\cite{Jacobson+2011b}\\
\end{table}
\end{landscape}
\twocolumn

\subsubsection{Final values from combined spectra}\label{final_rv}

The final values of the radial velocity for each star are derived from the combined 
spectra. For the cases of stars observed with
several instruments we adopt the weighted mean of all the determinations, and the mean of the nominal errors
as the uncertainty. These final values are found in column 12 of Table~\ref{lit}.

In general, stars have compatible radial 
velocities within the same cluster. This is because they were already pre-selected to be
very likely cluster members, as explained in Sec.~\ref{sec: sample}.
However, a re-analysis of membership is performed. We flag as non-members those stars which have $v_{\text{r}}$
not compatible at $3\sigma$ level of the radial velocity of the cluster.
We have used the median and the mean absolute deviation ($MAD$).
We iterate this by rejecting the non-members and recalculating the median
radial velocity, until we find a sample of compatible stars. Under this criterium we flag the following five stars:

\begin{enumerate}[(i)]
 \item NGC~1907 W2087 has a significant difference of $\sim60\, \text{km s}^{-1}$ with 
respect to the other stars from the
same cluster. The four values from individual 
exposures of this star (see Table~\ref{vrindiv})
are compatible with each other, so probably it is a non-member star or a large 
period spectroscopic binary. There is no
other measurement in the literature for comparison.
 \item NGC~2539 W233 has a radial velocity of $34.8\pm1.1\, \text{km s}^{-1}$, which is
$5.4\, \text{km s}^{-1}$ above the median of the other five stars. It was already flagged as spectroscopic binary by
\citet{Mermilliod+2008}. They obtain a variability with the maximum at 28.3$\pm$1.1 km s$^{-1}$. This value is compatible with ours within 3$\sigma$.
 \item NGC~2682 W224 has a radial velocity $6.5\, \text{km s}^{-1}$ under the median of
the cluster. The four individual spectra were taken in two consecutive days
and the individual radial velocities are in agreement. It was already flagged 
as member spectroscopic binary by \citet{Jacobson+2011b} and \citet{Geller+2015}.
 \item NGC~6819 W983 has a variable radial velocity as 
shown in Table~\ref{vrindiv} and discussed in Sec.~\ref{sec: rv indiv}. For this reason we do not give a 
final value of the radial velocity, and we do not include it in Table~\ref{lit}.
Neither \citet{Hole+2009} nor \citet{Milliman+2014} identify this star as a radial velocity variable, obtaining a final radial velocity of 
$2.36\pm 0.20$ km s$^{-1}$. Both studies are based in the same spectra (6 observations) and classify this star as single sember for having
e/i<4 (external error divided by internal error).
If this star was confirmed to be a
cluster binary member, we could consider it in the abundance analysis of the cluster.
 \item NGC~7762 W0084 has a large difference of $\sim$~$40\, \text{km s}^{-1}$ with respect to the other stars from 
the same cluster. Radial velocities obtained from the three
individual spectra acquired in two consecutive nights are consistent within the uncertainties.
There are neither previous radial velocity measurements nor information on membership for this cluster.
\end{enumerate}

Special attention must be payed to NGC~7789. Following the iterative procedure described above,
two stars should be rejected: W08260 and W07714. Radial velocities of all stars in this OC
compare well with the literature for stars in common \citep[][see Table~\ref{lit}]{Gim+1998,Jacobson+2011b},
which considers all of them as members.
Moreover, \citet{Jacobson+2011b} reported that they find a broader dispersion compared with other OCs.
Taking into account the OC mean radial velocity and dispersion from the three
large samples in the literature (Table~\ref{rv_mean}), all the seven stars studied here fall inside the distribution.
Therefore, we have decided to keep these two stars as members.

The rest of studied stars from the observed clusters are compatible
with being members of their parent cluster. We point out that stars NGC~1907 W0133,
NGC~6819 W978, and NGC~7762 W0003, have radial velocities outside of the $3MAD$ margin of the cluster, but when also
considering the uncertainties on these radial velocities, these stars are still within the cluster 
distributions, and are included as members in our sample
(see Fig.~\ref{meanRV}). The doubtful cases of membership will be probably solved when doing the abundance analysis.

\subsubsection{Radial velocities of clusters}\label{sec: clusters}
\begin{figure}
 \centering
\includegraphics[width=0.5\textwidth]{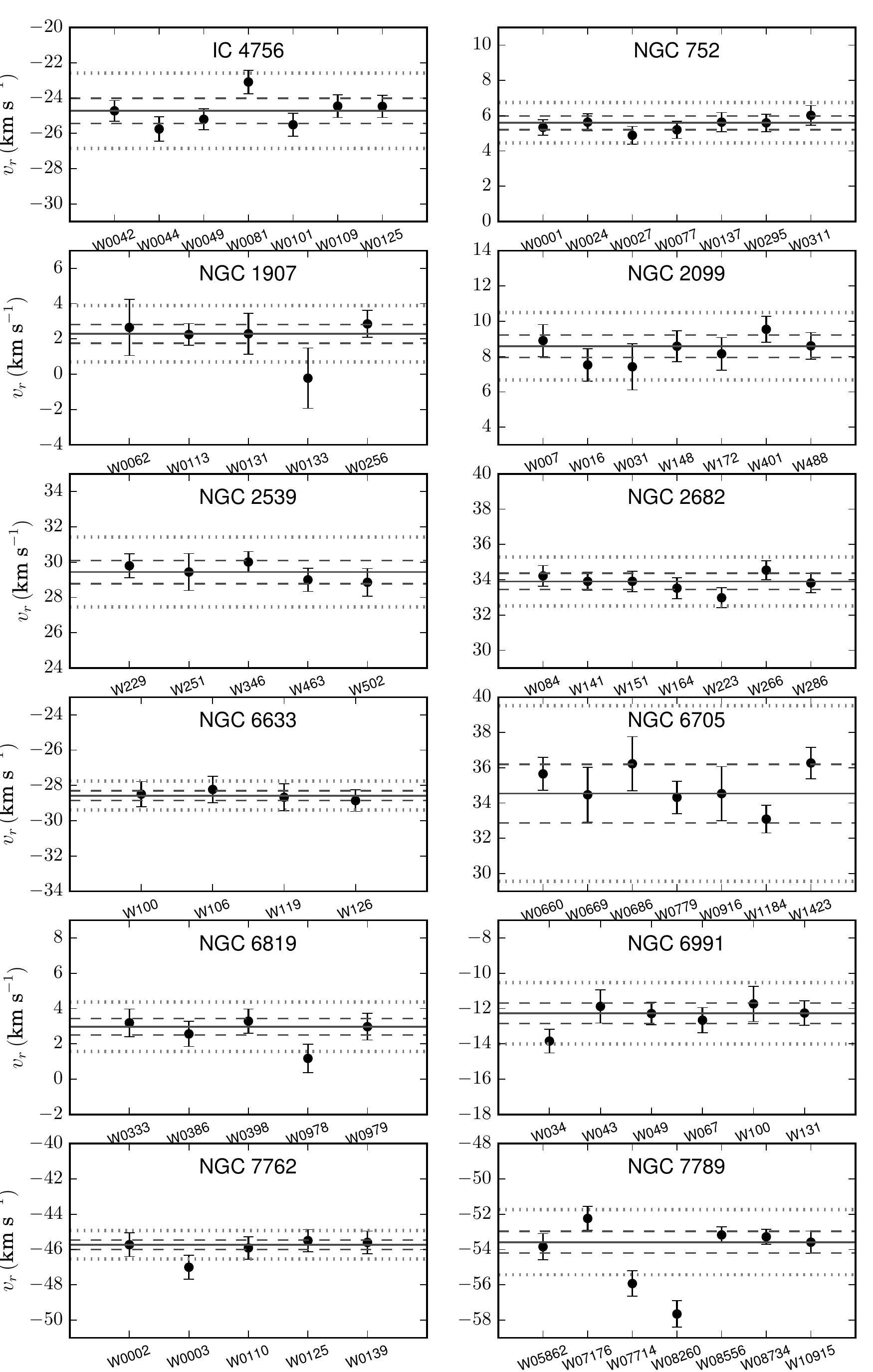}
 \caption{Radial velocities of the cluster stars considered as members.
The solid line corresponds to the median radial velocity of the cluster 
(calculated with the considered member stars), the dashed line corresponds to the mean absolute
deviation level $1MAD$, and the dotted line shows the $3MAD$ level.}\label{meanRV}
\end{figure}

The sample of non-spectroscopic binaries and bona-fide member stars is used to compute the cluster radial velocity. 
Median values and $MAD$ are found in Table~\ref{rv_mean} and plotted in Fig.~\ref{meanRV}. We also list in Table~\ref{rv_mean}
previous determinations of the cluster radial velocity, for those references where a mean value is given.
All values from literature are compatible within 3$\sigma$ with the ones derived here.

\begin{figure*}
\centering
\begin{minipage}{140mm}
\includegraphics[width=0.8\textwidth]{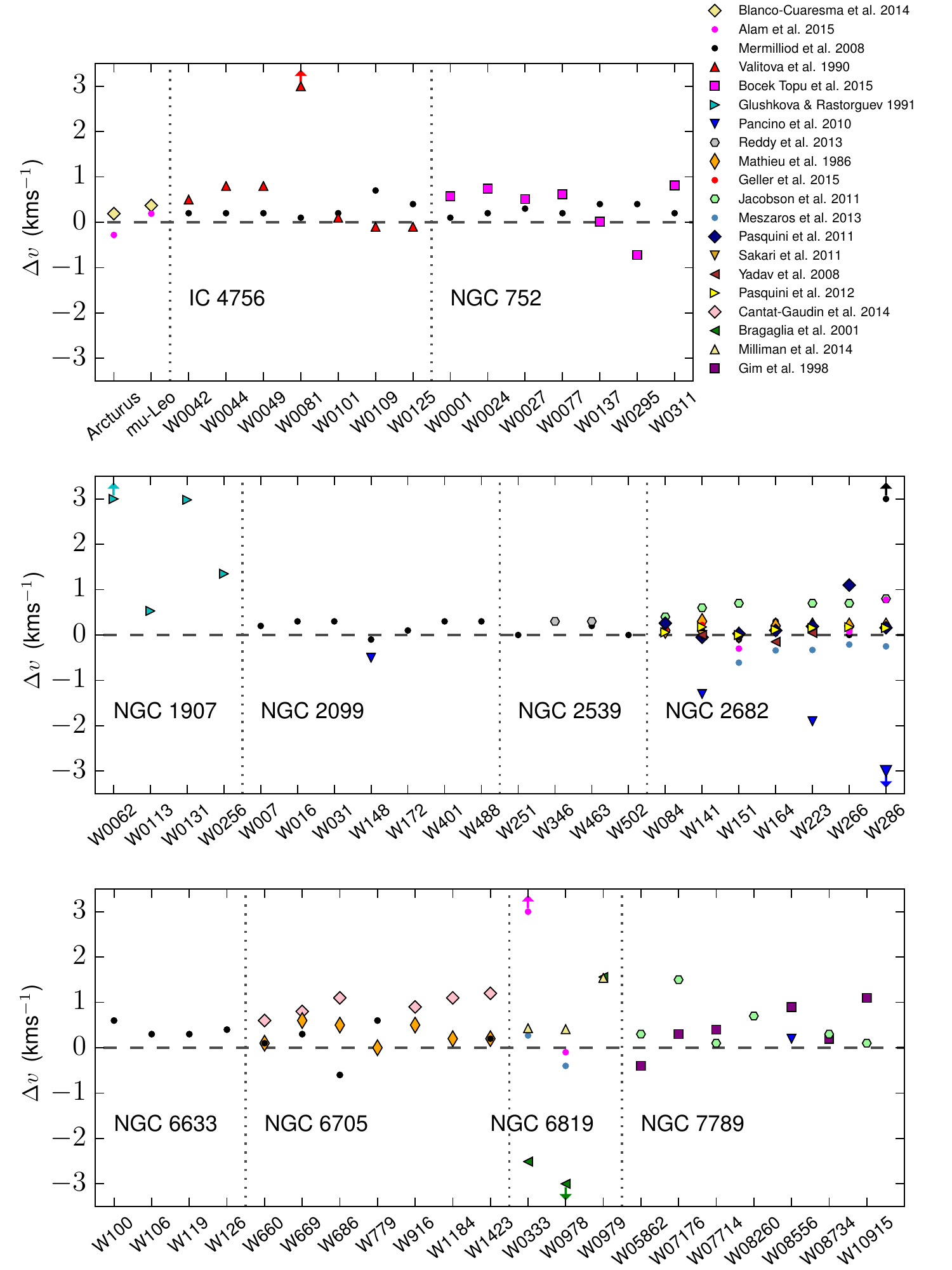}
\caption{\label{fig_lit}Radial velocity comparison with literature. Stars are
grouped by cluster. Differences, in the direction OCCASO-literature, are plotted 
for each star. Different points in the same x-coordinate denote different literature values for the 
same star. Points out of the
set y-limits are marked with an arrow. We have not plotted here stars NGC~2539 W233,
and NGC~2682 W224, for being possible spectroscopic binaries as explained in 
Sec.~\ref{final_rv}. Uncertainties are specified in Table~\ref{lit}.}
\end{minipage}
\end{figure*}

The radial velocity dispersions within each cluster are found between
$0.3-1.7\, \text{km s}^{-1}$. 
The quoted dispersions are the result of (a) the precision that we have in our radial velocity determinations
(Table~\ref{lit}), which is computed as the line-by-line
radial velocity variance found by DAOSPEC, (b) a fraction of undetected binaries,
and (c) the intrinsic internal dispersion of each cluster.
In most of the cases the dispersions in Table~\ref{rv_mean} are at the level of the quoted precisions.
Only, the dispersion for NGC~6705 is very well above the uncertainties ($1.7\, \text{km s}^{-1}$). 
This can be indicative that either this cluster has a larger fraction of undetected binaries, or
that this is indeed the instrinsic radial velocity dispersion, and that this OC is
kinematically hot. Given that the star by star comparison of this cluster with the literature is coherent
within the uncertainties (Fig.~\ref{fig_lit}, Table~\ref{lit}), 
we tend to think that this is the intrinsic
velocity dispersion. Moreover, this OC is the most massive
and youngest cluster in the sample. \citet{cantat-gaudin+2014} selected bona-fide members and found
a mean radial velocity of $34.1\pm 1.5\, \text{km s}^{-1}$ from 21 stars (UVES targets), and 
$35.9\pm 2.8\, \text{km s}^{-1}$ from 536 stars (GIRAFFE targets).
Our result confirms the high intrinsic velocity dispersion of this cluster.

\begin{table*}
 \begin{centering}
  \caption{Radial velocities of each cluster calculated as the median of the 
non-spectroscopic binaries and bona-fide member stars. The $MAD$ is assigned as the uncertainty, the
number of stars considered as members and used to derive the cluster radial velocity are written in parentheses.
Other determinations of the cluster radial velocity are shown in column 3, and the reference
is listed in column 4. Difference between OCCASO and literature is computed as 
$\Delta v=v_{\text{r}}-v_{\text{r,lit}}$. Notice: larger differences in the comparisons of NGC~1907 with
\citet{Glushkova+1991}, and NGC~7789 are commented in the main text (Sec.~\ref{sec: comparison} and \ref{final_rv}.)
}
\label{rv_mean}
  \begin{tabular}{@{}ccccc@{}}
 \hline
   Cluster & $v_{\text{r}} (\text{km s}^{-1})$ & $v_{\text{r,lit}} (\text{km s}^{-1})$ & $\Delta v_{\text{r,lit}} (\text{km s}^{-1})$ & Reference\\
 \hline
  IC~4756 & $-24.7\pm0.7$ (7) & $-25.0\pm 0.2$ (15) & 0.3& \citet{Valitova+1990}\\
 & & $-25.15\pm 0.17$ (17) & 0.45& \citet{Mermilliod+2008}\\
  NGC~752 & $5.6\pm 0.4$ (7)  & $5.04\pm 0.08$ (16)&0.56 &\citet{Mermilliod+2008}\\
 & & $4.82\pm 0.20$ (10) &0.78 & \citet{BocekTopcu+2015}\\
  NGC~1907 & $2.3\pm 0.5$ (5) & $0.1\pm 1.8$ (4) & 2.2&\citet{Glushkova+1991} \\
  NGC~2099 & $8.6\pm0.6$ (7)  & $8.30\pm 0.20$ (30) & 0.3& \citet{Mermilliod+2008} \\
  NGC~2539 & $29.4\pm0.7$ (5) & $28.89\pm 0.21$ (11) & 0.51&\citet{Mermilliod+2008}\\
  NGC~2682 & $33.9\pm0.5$ (7) & $33.52\pm 0.29$ (23) & 0.38& \citet{Mermilliod+2008} \\
  & & $33.73\pm 0.83$ (110) & 0.17&\citet{Pasquini+2011}\\
  & & $33.3\pm 0.6$ (22) & 0.6& \citet{Jacobson+2011b}\\
  & & $33.67\pm 0.09$ (141)&0.23 & \citet{Yadav+2008}\\
  & & $33.74\pm 0.12$ (77) & 0.16 &\citet{Pasquini+2012}\\
  NGC~6633 & $-28.6\pm0.3$ (4)& $-28.95\pm  0.09$ (6) & 0.35&\citet{Mermilliod+2008}\\
  NGC~6705 & $34.5\pm1.7$ (7) & $35.08\pm  0.32$ (15) & -0.58& \citet{Mermilliod+2008}\\
  & & $34.1\pm 1.5$ (21) & 0.4& \citet{cantat-gaudin+2014}\\
  NGC~6819 & $3.0\pm0.5$ (5)  & $2.45\pm1.02$ (566) & 0.55& \citet{Milliman+2014}\\
  NGC~6991 & $-12.3\pm0.6$ (6)& - & - \\
  NGC~7762 & $-45.7\pm0.3$ (5)& - & - \\
  NGC~7789 & $-53.6\pm0.6$ (7)& $-54.9\pm 0.9$ (50)& 1.3& \citet{Gim+1998}\\
  & & $-54.7\pm 1.3$ (26)& 1.1& \citet{Jacobson+2011b}\\
  & & $-54.6\pm 1.0$ (29) & 1.0& \citet{overbeek+2015} \\
 \hline
\end{tabular}
\end{centering}
\end{table*}

\subsection{Comparison with literature}\label{sec: comparison}

We compared our final values for each star (column 12 of Table~\ref{lit}),
with previous measurements in the literature, when available (column 13 of Table~\ref{lit}).
Since in most cases our individual exposures are taken during the same night,
this external comparison is also useful to identify potential spectroscopic binaries.

Calculated differences with each author are shown in Table~\ref{lit} (column 14) and illustrated in 
Fig.~7. We exclude from this comparison the
confirmed spectroscopic binaries already described in Sec.~\ref{final_rv} 
(NGC~6819 W983, NGC~2539 W233 and NGC~2682 W224).
The mean differences with each author are shown in Table~\ref{offset_lit}.

We find good agreement with literature except for five stars:
\begin{enumerate}[(i)]
 \item IC~4756 W0081: we find a difference of 4.8 km s$^{-1}$ with 
\citet{Valitova+1990}, and a difference of only
0.1 km s$^{-1}$ with \citet{Mermilliod+2008}. Given the small differences of the 
other stars in common with \citet{Valitova+1990}, we consider this case an outlier in this comparison and we exclude it to 
calculate the mean difference with these authors (Table~\ref{offset_lit}). Our three individual measurements are
taken within the same night (Table~\ref{vrindiv}), so we cannot know if this star is a spectroscopic binary. A large
set of measurements from \citet{Mermilliod+2008} do not show variability.
 \item NGC~1907 W0062: we find a difference of 4.68 km s$^{-1}$ with 
\citet{Glushkova+1991}. We have three other stars from the cluster NGC~1907
in common with these authors, with differences of: 0.53, 2.98, 1.35 km s$^{-1}$. Their
uncertainties are of the order of $1$ km s$^{-1}$. The mean 
difference with these authors is large
($2.4\pm 1.6$ km s$^{-1}$), even if we consider the star W0062 as outlier 
($1.6\pm1.0$ km s$^{-1}$). \citet{Glushkova+1991} reported large uncertainties
in their final values due to large errors in the observational data.
 \item NGC~6819 W0333: there is a discrepancy of -2.11 km s$^{-1}$ with 
\citet{Bragaglia+2001}, of 0.43 km s$^{-1}$ with \citet{Milliman+2014}, and $8.8\, \text{km s}^{-1}$ with
\citet{Alam+2015}, which is the Data Release 12 (DR12) of APOGEE. We find a difference of only $0.7\, 
\text{km s}^{-1}$ with \citet{Meszaros+2013}, which is the Data Release 10 (DR10).
This star is reported to have 
``high persistency''\footnote{The APOGEE detector suffers of the persistence effect, where
the amount of charge deposited can be affected by the previous exposure.
This is further explained in \citet{Nidever+2015}.} in the APOGEE detector by \citet{Alam+2015}.
Given the low differences of the other stars in common, this 
effect could be the explanation for the discrepancy. From a set of 5 measurements \citet{Milliman+2014} identify this star as single member.
 \item NGC~6819 W0978: there is a difference of $-4.76\, \text{km s}^{-1}$ with 
\citet{Bragaglia+2001}, and a small difference with both
APOGEE DR10 and DR12, -0.4 and -0.1 km s$^{-1}$, respectively. Also we see a small difference of
0.41 km s$^{-1}$ with \citet{Milliman+2014}, which identify this star as single member.
\cite{Bragaglia+2001} have used a spectral resolution of $R=40,000$. They do not
specify their errors, but they report that they were not interested in obtaining precise radial velocities.
 \item NGC~2682 W286 we find significant differences of 8.1 km 
s$^{-1}$ and -5.1 km s$^{-1}$ with \citet{Mermilliod+2008} and \citet{Pancino+2010}, respectively. Since
we find differences smaller than 1 km s$^{-1}$ for the same star with six other 
autors \citep{Pasquini+2011, Jacobson+2011b, Pasquini+2012,
Alam+2015, Meszaros+2013, Mathieu+1986}, we consider this case as 
outlier, and we exclude it to calculate the mean difference
with \citet{Mermilliod+2008} and \citet{Pancino+2010} in Table~\ref{offset_lit}.
\end{enumerate}

We can state that large differences are found for few specific authors and stars.
Given that for the same stars we find compatible values with other authors, we do not interpret these
discrepancies as due to binarity but some spurious measurements in the literature.
For all these stars mentioned above we make use of our radial velocities.

Arcturus and $\mu$-Leo are compared with the values given
by \citet{Blanco+2014} for the \textit{Gaia} Benchmark stars. These are two stars with 
very precise determination of the radial velocity
because they are taken as standard stars for the \textit{Gaia} mission wavelength 
calibration. We find a difference of 0.19 and 0.37
km s$^{-1}$, respectively. We also compare with the results
for the APOGEE DR12, which are -0.28 and 0.19 km s$^{-1}$, respectively.
All differences are lower than our quoted uncertainties.

We compare the 6 stars in common with GES for the cluster NGC~6705 with 
\citet{cantat-gaudin+2014} (21 stars analyzed), finding a 
mean offset of $0.95\pm0.21 \,\text{km s}^{-1}$. However, comparison of individual stars agree within
the quoted uncertainties.

Besides, we have 7 stars in common with APOGEE DR12 \citep{Alam+2015}, and 8 stars in 
common with APOGEE DR10 \citep{Meszaros+2013}. To make an overall comparison we do not
take into account the star NGC~2682 W224 and NGC~6819 W0333 for the reasons already discussed.
We find a mean offset of $0.06\pm0.34 \,\text{km s}^{-1}$ with \citet{Alam+2015}, and
$-0.27\pm0.25 \,\text{km s}^{-1}$ with \citet{Meszaros+2013}.

All the computed mean differences with literature estimates are listed in Table~\ref{offset_lit}. 
The largest offset is found for \citet{Glushkova+1991} and is already commented above.
The mean of the differences with the other authors is $0.2\pm 0.7$ km s$^{-1}$. This means that the
accuracy with the overall literature is formally consistent with the quoted uncertainties.

\begin{table}
 \begin{centering}
  \caption{Mean offsets and dispersions calculated for each author from the
values in Table~\ref{lit}. Offsets (second column) are
in the direction OCCASO-literature, the number of stars for each paper is 
listed in the third column.\label{offset_lit}}
  \begin{tabular}{@{}ccc@{}}
 \hline
   Reference & $\Delta v_{\text{r}} (\text{km s}^{-1})$ & $N$ \\
 \hline
\citet{Blanco+2014} & $0.28\pm0.09$ & 2 \\
\citet{Mermilliod+2008}$^1$ & $0.21\pm0.21$ & 40\\
\citet{Valitova+1990}$^2$ & $0.33\pm0.39$ & 6\\
\citet{Glushkova+1991} & $2.4\pm1.6$ & 4\\
\citet{Pancino+2010}$^3$ & $-0.88\pm 0.79$ & 4\\
\citet{cantat-gaudin+2014} & $0.95\pm0.21$ & 6\\
\citet{Mathieu+1986} & $0.24\pm0.18$ & 14\\
\citet{Bragaglia+2001} & $-0.5\pm2.0$ & 2\\
\citet{Gim+1998} & $0.42\pm0.49$ &6 \\
\citet{Alam+2015}$^4$ & $0.06\pm0.34$ & 7\\
\citet{Meszaros+2013} & $-0.27\pm0.25$ &7 \\
\citet{Pasquini+2011} & $0.26\pm0.36$ & 7\\
\citet{Sakari+2011} & $0.00$ & 1 \\
\citet{Yadav+2008} & $-0.05\pm0.07$ & 3\\
\citet{Pasquini+2012} & $0.12\pm 0.06$ & 7\\
\citet{Milliman+2014} & $0.13\pm 0.06$ & 3\\
\citet{BocekTopcu+2015} & $0.4\pm 0.5$ &7\\
\citet{Geller+2015} & $0.4\pm 0.5$ & 5\\
 \hline
\end{tabular}
\end{centering}
\\

$^1$excluded NGC~2682 W286\\
$^2$excluded IC~4756 W0081\\
$^3$excluded NGC~2682 W286\\
$^4$excluded NGC~6819 W0333
\end{table}

\subsection{Discussion: relation to the disc kinematics}\label{sec: discussion}
As described in Sec.~\ref{sec: science}, Galactic disc kinematics is one of the science topics of
OCCASO. This section is devoted to a preliminary analysis with the 12 OCs published here.
A more detailed investigation will be carried out when all observations will be completed and 
{\em Gaia} proper motions will be available. Our analysis here is also limited by the small range 
of Galactocentric distances of the 12 OCs, mainly in the range 8--10 kpc.
Most of the OCs studied here are located in the vicinity of the Local arm.
Three of them in the Perseus arm, and only NGC~6705, is located in the Sagittarius arm (see Fig.~\ref{fig:cinematica}).

\subsubsection{Radial velocity with respect to the GSR and RSR}

It is well known that the Galactocentric velocity of any source in the Galactic disc can be described using two
components: (a) the velocity associated to a circular orbit around the Galactic center, constrained 
by the Galactocentric distance and defining the Regional Standard of Rest (RSR), and (b) an additional
peculiar velocity, the velocity with respect to such RSR. The velocity with respect to RSR tells us
how much the motion of the cluster differs from the Galactic disc rotation.  

One can compute the velocity with respect to the Galactocentric Standard of Rest (GSR) by adding
the spatial velocity of the Sun to the measured heliocentric velocity. This spatial velocity of the Sun is described in the
same two components: its velocity with respect to the Local Standard of Rest (LSR), and the circular motion of the LSR.
Considering only the line-of-sight component:

\begin{align}
 v_\text{GSR}=v_\text{r}+U_{\odot}\cos l \cos b + \left( \Theta_{\text{0}}+V_{\odot} \right) \sin l \cos b + W_{\odot} \sin b
\end{align}
where $v_r$ is the heliocentric radial velocity,  ($U_{\odot}$, $V_{\odot}$, $W_{\odot}$)
are the components of the motion of the Sun with respect to the LSR, and $\Theta_0$ is the circular velocity
at the Galactocentric distance of the Sun $R_0$.

The line-of-sight velocity with respect to the RSR can be computed by subtracting the circular motion 
of the RSR projected onto the line-of-sight:
\begin{align}
 v_\text{RSR}=v_\text{GSR}-  \Theta_R \frac{R_0}{R} \sin l \cos b 
\end{align}
where $\Theta_R$ is the circular velocity at the Galactocentric distance of the cluster $R$.
In first order aproximation (enough for the $R$ of our clusters) $\Theta_R$ is computed as
\begin{align}
\Theta_R=\Theta_{\text{0}}+\frac{\text{d} \Theta}{\text{d} R}(R-R_{0})
\end{align}

\noindent

Assuming the Sun motion derived by \citet{Reid+2014}\footnote{Values obtained by their model A5.} 
($U_{\odot}$, $V_{\odot}$, $W_{\odot}$)=(10.7,15.6,8.9) $\text{km s}^{-1}$, and their values of the Galactic 
rotation curve $\Theta_{\text{0}}=240\, \text{km s}^{-1}$, $R_{0}=8.34\, \text{kpc}$ and 
$\nicefrac{\text{d} \Theta}{\text{d} R}=-0.2\, \text{km s}^{-1}$, we derive $v_\text{GSR}$ and $v_\text{RSR}$
for each cluster. 
Galactocentric distances $R$ are computed from heliocentric distances in
\citet{Dias+2002}\footnote{Available at http://irsa.ipac.caltech.edu} (see Table~\ref{clusters}).
Since no error estimates are given for those distances, we adopted an uncertainty
of 0.2 mag in distance modulus, rather typical when determining distances from isochrone fitting. 
The errors in $v_\text{GSR}$ are computed taking into account errors in $v_\text{r}$, and the motion of the Sun: $\Theta_{\text{0}}$, 
$U_{\odot}$, $V_{\odot}$, and $W_{\odot}$. The errors in $v_\text{RSR}$ are computed taking into account also the errors in distance modulus. 

\begin{figure}
\centering
\includegraphics[width=0.5\textwidth]{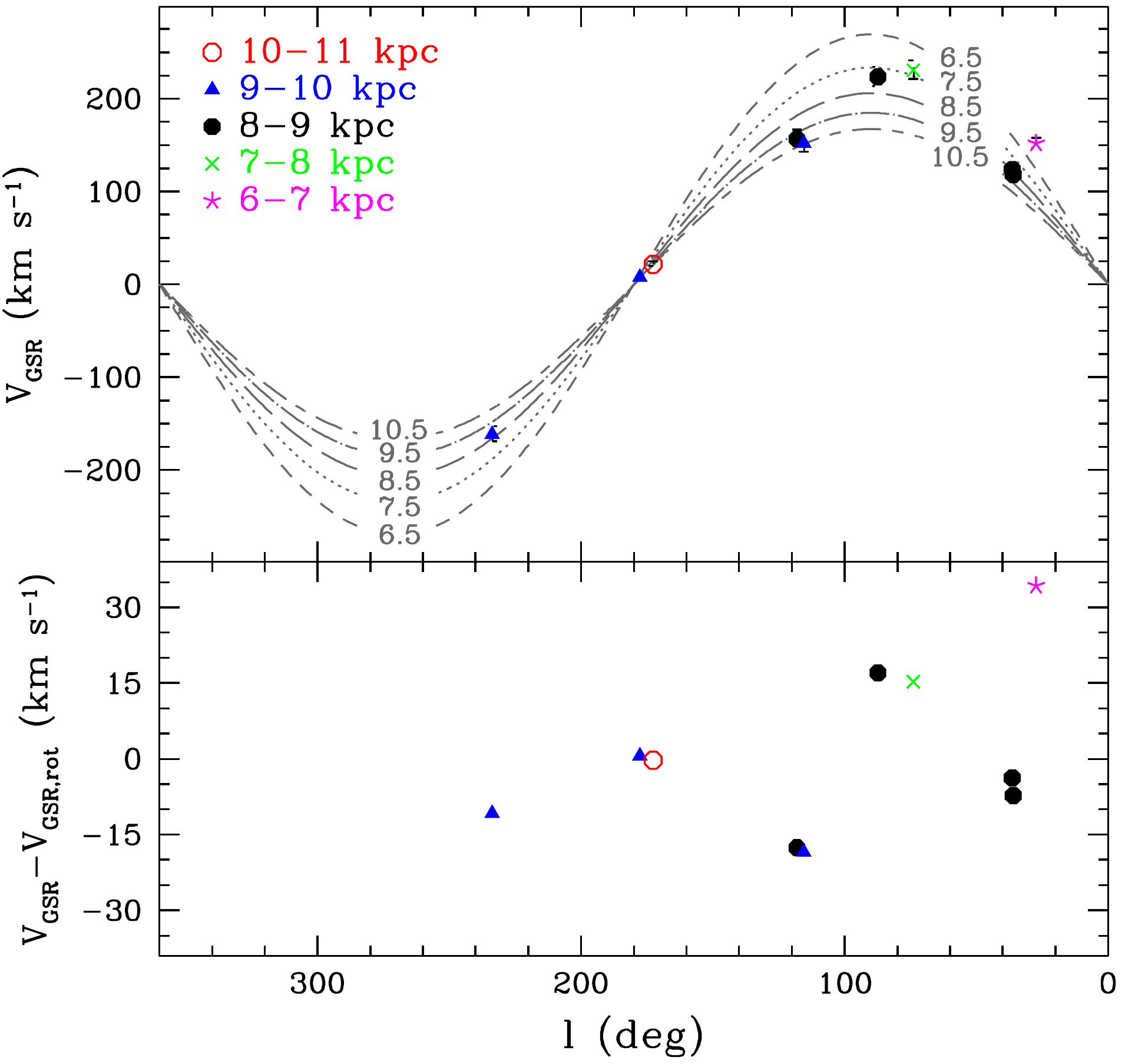}
\caption{Upper pannel: distribution of the studied clusters in the $l-v_{\text{GSR}}$ plane. The symbols change
as a function of Galactocentric radius. $v_{\text{GSR}}$ has been computed assuming ($U_{\odot}$, $V_{\odot}$, $W_{\odot}$)=(10.7,15.6,8.9)
$\text{km s}^{-1}$ and $\Theta_{\text{0}}$=240 km s$^{-1}$ from \citet{Reid+2014}. Lines represent circular orbits at different radii
showing the rotation curve derived by \citet{Reid+2014}. Errors in $v_{\text{GSR}}$ are not plotted since they are smaller
than the point size (see Table~\ref{rv_galac}). Lower pannel: differences between the velocities of the clusters with respect to the GSR 
and the circular velocity at the position of each cluster $v_{\text{GSR}}-v_{\text{GSR,rot}}$.}
\label{fig:cinematica_rot}
\end{figure}

Figure~\ref{fig:cinematica_rot} presents $v_\text{GSR}$ as a function of Galactic longitude\footnote{OCs at $b>15\deg$ (NGC~2682 and
NGC~752) are not plotted since at these latitudes the line-of-sight component of the velocity is not in the Galactic plane.}. The values 
corresponding to circular orbits at different radii have been overplotted.
There is a good correlation
between the Galactocentric distance of each cluster and the corresponding circular orbits, meaning that 
line-of-sight $v_\text{RSR}$ are small. The obtained values of $v_\text{RSR}$ and $v_\text{GSR}$ are listed in Table~\ref{rv_galac}.
The $v_\text{RSR}$ are in the range of $-27$ to $+24.7\, \text{km s}^{-1}$, typical values for the disc populations.
Mean $v_\text{RSR}$ of the eight clusters located in the Local arm is $-2\,\text{km s}^{-1}$ with an
standard deviation of $14\,\text{km s}^{-1}$. Again, rather typical.

We have also computed $v_\text{RSR}$ using different 
assumptions for the Galactic rotation and Sun's location taken from \citet{Antoja+2011} and \citet{Sofue+2009}.
The mean differences of $v_\text{RSR}$ from the different assumptions are smaller than $0.4\,\text{km s}^{-1}$, 
well within uncertainties due to the errors in radial velocity and distances. 
Therefore, our $v_\text{RSR}$ do not favour one or another Galactic rotation curve or location of the Sun.

\begin{table}
 \centering
  \caption{Radial projections of the velocities with respect to the Regional Standard of Rest $v_{\text{RSR}}$, and the Galactic
Standard of Rest $v_\text{GSR}$.\label{rv_galac}}
  \begin{tabular}{@{}ccc@{}}
 \hline
   Cluster & $v_\text{GSR}$ & $v_\text{RSR}$ \\
& (km s$^{-1}$) & (km s$^{-1}$)\\
 \hline
\multicolumn{2}{l}{\it Saggitarius arm:} \\
NGC 6705 &  $151.9 \pm 5.2$ & $ 24.7^{+  3.3}_{  -2.9}$\\
\\
\multicolumn{2}{l}{\it Local arm:} \\
IC 4756  &  $123.8 \pm 6.4 $ & $-13.1^{+  1.0}_{  -1.1}$\\
NGC 752  &  $142.1 \pm 6.7 $ & $ 10.7^{+  0.6}_{  -0.6}$\\
NGC 2539 &  $-161.4 \pm 8.4 $& $ -5.6^{+  1.7}_{  -1.6}$\\
NGC 2682 &  $-85.6 \pm 5.4 $ & $ 14.8^{+  0.9}_{  -0.8}$\\
NGC 6633 &  $118.3 \pm 6.3 $ & $-15.2^{+  0.5}_{  -0.6}$\\
NGC 6819 &  $230.8 \pm 10.0 $& $ 12.2^{+  0.8}_{  -0.8}$\\
NGC 6991 &  $223.6 \pm 10.5 $& $  3.8^{+  0.7}_{  -0.7}$\\
NGC 7762 &  $156.7 \pm 9.3 $ & $-27.0^{+  1.0}_{  -1.1}$\\
\\
\multicolumn{2}{l}{\it Perseus arm:} \\
NGC 1907 &  $21.7 \pm 2.5 $  & $ -1.2^{+  1.2}_{ -1.2}$\\
NGC 2099 &  $7.9 \pm 2.0 $   & $  0.2^{+  0.7}_{ -0.7}$\\
NGC 7789 &  $152.0 \pm 9.6 $ & $-24.0^{+  2.6}_{ -2.7}$\\
 \hline
\end{tabular}
\end{table}

\subsubsection{Spatial velocity with respect to RSR}

Cluster line-of-sight velocities were combined with proper motions to derive full spatial velocities. To do so,
mean proper motions were taken from \citet{Dias+2014} and are listed in Table~\ref{table:pm}. 
\citet{Dias+2014} compared their mean proper motions with other values in the literature and concluded that 
mean differences and standard deviation were among 1.4--1.7 mas yr$^{-1}$.
We have assumed uncertainties of 1.5 mas yr$^{-1}$ in each proper motion coordinate. The velocity with
respect to RSR in a cartesian Galactocentric frame, ($U_s$, $V_s$, $W_s$), was computed as (more details in the derivation
in \citet{Reid+2014}):
{\scriptsize
\begin{align}
\left( \begin{array}{c}
U_s\\
V_s+\Theta_R\\
W_s\\ \end{array} \right) = \text{R}_z\left(-\beta\right) \left[\left( \begin{array}{c}
U_{\odot}\\
V_{\odot}+\Theta_{0}\\
W_{\odot}\\ \end{array} \right) + \text{R}_z\left(-l\right) \text{R}_y\left(b\right)\left( \begin{array}{c}
v_r\\
D \mu_l \cos b\\
D \mu_b\\ \end{array} \right) \right]
\end{align}
}
\noindent where $U_s$ points towards the Galactic Center, $V_s$ towards Galactic rotation, and $W_s$ towards the North
Galactic Pole, $\text{R}_z$ and $\text{R}_y$ are rotations of a certain angle on the z and y axis respectively,
$\beta$ is the angle formed by Sun - Galactic Center - Cluster, $\mu_l$ and $\mu_b$ are the proper motions in the $l$, $b$ directions.

The uncertainty has been derived from 
classical Markov chain Montecarlo simulation with 10\,000 random realizations for each cluster.

Taking the values from Table~\ref{table:pm} we find mean values and standard deviations of 
$\langle U_s\rangle=-6\pm15\, \text{km s}^{-1}$, $\langle V_s\rangle=-9\pm24\, \text{km s}^{-1}$,
$\langle W_s\rangle=7\pm23\, \text{km s}^{-1}$.
Studies of velocity dispersions as a function of age such as \citet[][fig. 7]{Holmberg+2009} indicate that 
for stars of ages 0.8-2.5 Gyr we expect $\sigma_U$ and $\sigma_V$ between 15-25 $\text{km s}^{-1}$.
So, this is well verified in our sample.
There are only four OCs, NGC6705, NGC6819, NGC7762 and NGC7789, with 
velocities with respect to their RSR larger than about $30\,\text{km s}^{-1}$ and are the ones with the
larger errors. Particularly remarkable is NGC6819 with a vertical velocity of $71.73\pm23.10 \, \text{km s}^{-1}$.

IC~4756 and NGC~6633, both in the Local arm, are located close together and have similar age and spatial
non-circular velocity. Taken together, this may indicate some relationship in their formation. 
Better uncertainties
in proper motions like the ones that \textit{Gaia} will provide, and comparison of chemical abundances,
(which is the main purpose of OCCASO) will clarify this issue.

\begin{figure}
\centering
\includegraphics[width=0.5\textwidth]{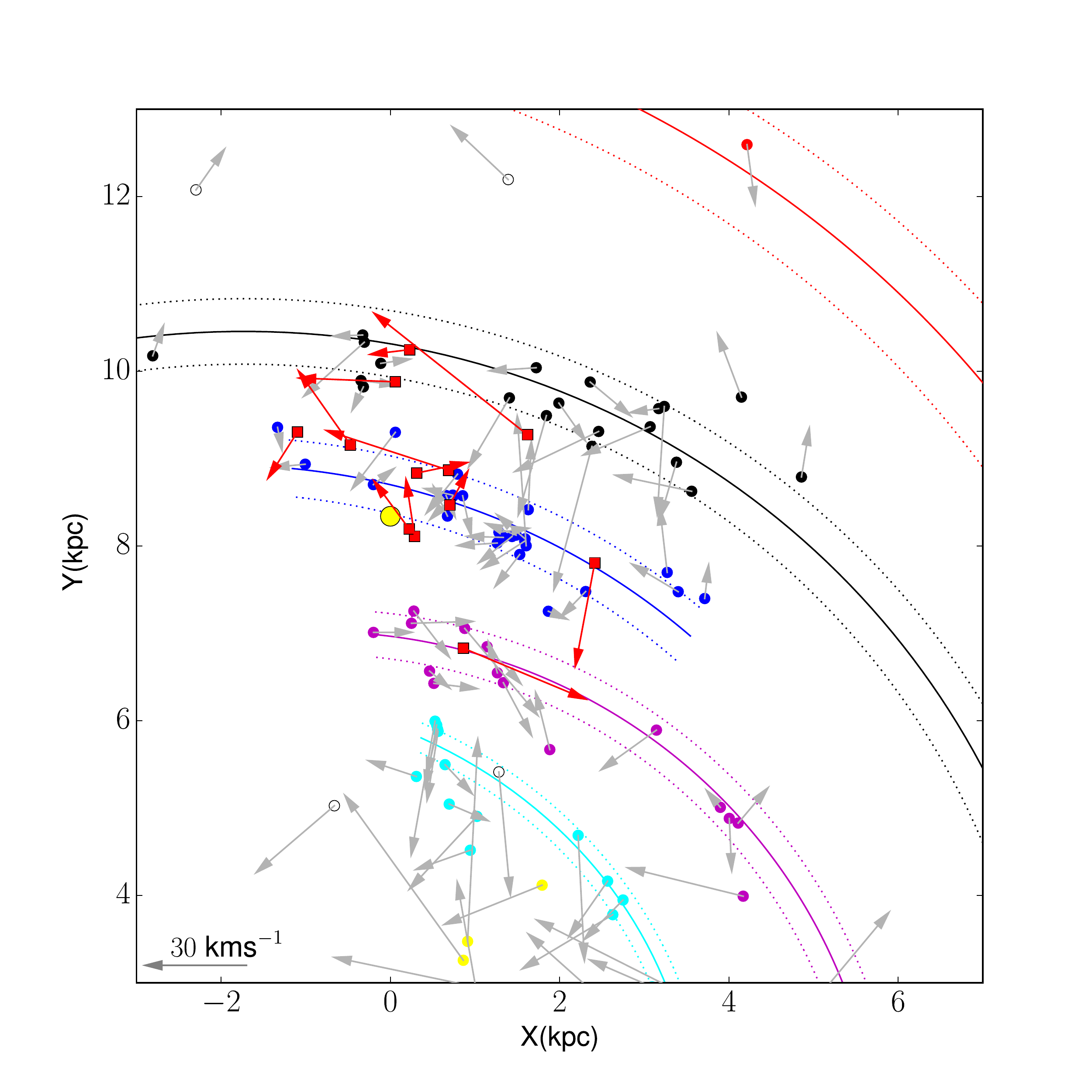}
\caption{Spatial distribution of the 12 studied clusters in this paper (red squares). 
The Sun (big yellow circle) is at (0,8.34) kpc. The Galaxy's
spiral arms positions and widths (coloured solid and dashed lines) are obtained from \citet{Reid+2014}.
Coloured circles show the locations of high-mass star forming regions studied in \citet{Reid+2014}.
Circles are coloured according to the spiral arm to which are assigned to, as described in \citet{Reid+2014}.
The arrows show the spatial velocity with respect to the RSR projected onto the plane from \citet{Reid+2014}
(in grey), and from this study (red).}
\label{fig:cinematica}
\end{figure}

Finally, in Fig.~\ref{fig:cinematica} we have plotted the spatial distribution of the 12 OCs in 
the Galactic plane. The location of the spiral arms, as derived by \citet{Reid+2014},
and the ($U_s$, $V_s$) components for each cluster have been overplotted. High-mass star forming regions
(HMSFR) studied by \citet{Reid+2014} are also included.
We have calculated mean values and dispersions of the HMSFR
$\langle U_s\rangle$, $\langle V_s\rangle$, $\langle W_s\rangle$ in each arm. And we have computed differences
between the OCs components and these mean values (see last three columns in Table~\ref{table:pm}), to see if
there exists a hint of dynamical relationship between our OCs and the arms. In general, the differences fall inside the $3\sigma$
margin except for the clusters NGC~7789 (Perseus arm), NGC~7762 and NGC~6819 (Local arm),
and NGC~6705 (Saggitarius arm). We do not find correlations with age, but our sample is limited in number. Again,
precise proper motions of \textit{Gaia} can help on the interpretation of the kinematics of the studied clusters.

\begin{table*}
\caption{$U_s$, $V_s$ and $W_s$ are the components of the
non-circular velocity at the position of each cluster. These are computed
from proper motions \citep{Dias+2014} and our radial velocities, using the values for the motion
of the Sun with respect to the LSR from \citet{Reid+2014}. Mean values and dispersions of the non-circular velocity
for the HMSRF studied by \citet{Reid+2014} are indicated for each arm in italics. The last three columns list the
 differences in the direction OCCASO - $\langle$HMSFR$\rangle$.}
\label{table:pm}
\begin{tabular}{lcccccccc}
\hline
Cluster       & $\mu_\alpha$cos$\delta$ &$\mu_\delta$ &  $U_s$ & $V_s$ & $W_s$ & $\Delta U_s$ & $\Delta V_s$ & $\Delta W_s$ \\
              &(mas yr$^{-1}$) & (mas yr$^{-1}$) & (km s$^{-1}$) & (km s$^{-1}$)& (km s$^{-1}$) & (km s$^{-1}$) & (km s$^{-1}$)& (km s$^{-1}$)\\
\hline
\multicolumn{3}{l}{\it Saggitarius arm:} & \emph{4.44$\pm$10.36} & \emph{3.98$\pm$11.51}& \emph{-3.60$\pm$6.75}& & &\\
NGC 6705 &  -1.23  &  1.31  &  3.77$\pm$ 7.83  & 39.10$\pm$11.14  & 22.44$\pm$13.68 & -0.67 & 35.11 &26.03\\
\\                                                                              
\multicolumn{3}{l}{\it Local arm:} & \emph{1.96$\pm$10.44}& \emph{-3.90$\pm$4.51}& \emph{5.01$\pm$10.16}& & &\\
IC 4756  &  -0.60  & -1.69  &-15.17$\pm$ 2.27  & -2.74$\pm$ 2.78  &  6.15$\pm$ 3.47 & -17.13 & 1.16 & 1.14 \\
NGC 752  &   1.81  & -3.90  & -4.25$\pm$ 2.99  & 12.17$\pm$ 3.16  &  0.47$\pm$ 3.13 & -6.21& 16.07 & -4.54\\
NGC 2539 &  -3.20  & -1.24  & 14.82$\pm$ 7.41  & -8.57$\pm$ 6.83  & -7.45$\pm$ 9.96 & 12.86& -4.67 &  -12.47\\
NGC 2682 &  -9.40  & -4.87  &-19.35$\pm$ 4.48  &-14.07$\pm$ 5.27  & -8.32$\pm$ 5.97 & -21.31& -10.17 & -13.33\\
NGC 6633 &  -2.27  & -4.95  &-12.75$\pm$ 1.77  & -9.16$\pm$ 2.26  &  4.42$\pm$ 2.67 & -14.71& -5.26 & -0.59\\
NGC 6819 &  -6.07  & -3.57  & 17.15$\pm$21.00  &  2.67$\pm$ 3.38  & 71.73$\pm$23.10 & 15.19& 6.57 &  66.72\\
NGC 6991 &  -1.50  &  1.94  &-10.46$\pm$ 6.09  &  3.04$\pm$ 0.65  & 18.40$\pm$ 6.59 & -12.42& 6.94 &  13.39\\
NGC 7762 &   3.44  & -2.21  &-10.45$\pm$12.84  &-38.51$\pm$ 7.91  &-10.35$\pm$ 6.31 & -14.41& -34.61 &  -15.36\\
\\
\multicolumn{3}{l}{\it Perseus arm:} & \emph{4.44$\pm$10.36}& \emph{3.98$\pm$11.51}& \emph{-3.60$\pm$6.75}& & &\\
NGC 1907 &  -0.85  & -4.22  & -0.07$\pm$ 1.53  & -7.98$\pm$13.63  &-17.89$\pm$14.94 & -4.51& -11.96 &  -14.29 \\
NGC 2099 &   2.08  & -6.40  & -1.37$\pm$ 0.91  &-28.28$\pm$11.24  &  1.75$\pm$11.22 & -5.81 & -32.26 & 5.35\\
NGC 7789 &   2.86  & -0.74  &-36.58$\pm$18.89  &-55.50$\pm$13.93  & -1.62$\pm$13.48 & -41.02& -59.48 &  1.98 \\
\hline
\end{tabular}
\end{table*}

\section{Summary}\label{sec:summary}
The OCCASO survey has been designed to obtain radial velocities and homogeneous abundances
for more than 20 chemical species for RC stars in a sample of 25 Northern OCs with ages $\gtrsim$0.3 Gyr.
These data will allow us to properly analyze the existence of trends with $R_{\text{GC}}$, $z$ and age,
in the Galactic disc. Moreover, our sample of OCs is complementary to GES-UVES observations of intermediate-age and old
Southern OCs. For this reason we include OCs in common with GES to guarantee homogenity between both surveys.
At the end of both surveys, an homogeneous sample of chemical abundances for around 50 OCs will be available.

We have collected observational data from high-resolution spectroscopy during 53 nights of observation
using the fiber-fed echelle spectrographs FIES and HERMES at the ORM, and CAFE at CAHA.
We have done a comparison among the results from the three instruments used, obtaining a
good agreement within the uncertainties.

The radial velocity analysis has been performed for 77 stars in 12 OCs.
We have derived radial velocities from 401 individual exposures. With these values we have found a new possible
spectroscopic binary NGC~6819 W983, which has never been identified as a multiple system.
We have derived radial velocities from the combined spectra with SNR$\geq 70$, obtaining uncertainties of $0.5-0.9\,\text{km s}^{-1}$.
We have used these values of the radial velocities to confirm or discard membership from our sample of stars
and compute a median radial velocity for each OC. In particular, we have obtained radial velocities
for OCs never studied before with high-resolution spectroscopy: NGC~1907 ($v_{r}=2.3\pm 0.5\,\text{km s}^{-1}$),
NGC~6991 ($v_{r}=-12.3\pm 0.6\,\text{km s}^{-1}$) and NGC~7762 ($v_{r}=-45.7\pm 0.3\,\text{km s}^{-1}$).

The radial velocities obtained in this paper agree with the values from previous
authors within the uncertainties, except for few cases. We have compared
the stars in common with other two large spectroscopic surveys: GES, 6 stars in common with an average difference
of $\Delta v_{r}=0.95\pm 0.21\,\text{km s}^{-1}$; and APOGEE, 7 stars in common with \citet[DR10]{Meszaros+2013} a mean difference
$\Delta v_{r}=-0.27\pm 0.25\,\text{km s}^{-1}$, and 7 stars in common with \citet[DR12]{Alam+2015} a mean difference of 
$\Delta v_{r}=0.06\pm 0.34\,\text{km s}^{-1}$.

Median radial velocities for each OC have been used to study their kinematics in relation to the disc and
the spiral arms. It is shown that all of the studied clusters follow the expected rotation of the Milky Way assuming
the rotation curve derived by \citet{Reid+2014}.

Adding information of proper motions
from \citet{Dias+2002} we have derived full spatial velocities, and we have compared the non-circular velocities among them.
There seems to be no clear relation of the peculiar velocities among the OCs from the same spiral arm (except for IC~4756 and NGC~6633), nor with
the peculiar velocities of the high-mass star-forming regions \citep{Reid+2014} from the same arms. From our sample
we calculate the dispersion in the two components of the plane velocity:
$\sigma_U$ and $\sigma_V=$ 15 and 24 $\,\text{km s}^{-1}$, which is expected for a population of ages
0.8--2.5 Gyr as seen in \citet{Holmberg+2009}.

\bibliographystyle{mn2e} 
\bibliography{biblio_v2} 

\section*{Acknowledgments}
We are greatful to the referee for the suggestions that improved this work.
We warmly thank F. Figueras for her useful comments and discussions.

This research made use of the WEBDA database, operated at the Department
of Theoretical Physics and Astrophysics of the Masaryk University, and the
SIMBAD database, operated at the CDS, Strasbourg, France.
This work was supported by the MINECO (Spanish Ministry of Economy) -
FEDER through grant ESP2013-48318-C2-1-R and ESP2014-55996-C2-1-R and
MDM-2014-0369 of ICCUB (Unidad de Excelencia 'Mar\'ia de Maeztu').

CG, CEM-V and RC acknowledge support from the IAC (grantP/301204) and from the Spanish
Ministry of Economy and Competitiveness (grant AYA2014-56795).

LC acknowledges financial support from the University of Barcelona under the APIF grant, and
the financial support by the European Science Foundation (ESF), in the framework of the GREAT Research Networking Programme.

\label{lastpage}

\end{document}